\documentclass[a4paper,onesided,12pt]{report}
\usepackage{styles/fbe_tez}
\usepackage[utf8x]{inputenc} % To use Unicode (e.g. Turkish) characters

\usepackage{amsmath, amsthm, amssymb}
\usepackage[bottom]{footmisc}
\usepackage{cite}
\usepackage{graphicx}
\usepackage{longtable}
\graphicspath{{figures/}} % Graphics will be here 

\usepackage{multirow}
\usepackage{subfigure}
\usepackage{algorithm}
\usepackage{algorithmic}
\usepackage{xcolor}

\usepackage{tabularx}
\usepackage{makecell}
\usepackage{tablefootnote}
\usepackage[hyphens]{url}
\newcolumntype{L}{>{$}l<{$}} % math-mode version of "l" column type
\newcolumntype{R}{>{$}r<{$}} % math-mode version of "l" column type
\usepackage[polutonikogreek,english]{babel}
\usepackage{notoccite}

\usepackage[a4paper,
            bindingoffset=0.2in,
            left=3.5cm,
            right=2cm,
            top=4cm,
            bottom=2cm,
            footskip=.25in]{geometry}

\newtheorem{definition}{Definition}[chapter]

\title{ACHIEVING ULTRA-RELIABLE LOW-LATENCY COMMUNICATION (URLLC) IN NEXT-GENERATION CELLULAR NETWORKS WITH PROGRAMMABLE DATA PLANES}
\turkcebaslik{PROGRAMLANABİLİR VERİ DÜZLEMLERİ İLE GELECEK NESİL HÜCRESEL AĞLARDA AŞIRI GÜVENİLİR DÜŞÜK GECİKMELİ İLETİŞİM'İN SAĞLANMASI (URLLC)}
\degree{B.S., Computer Engineering, Boğaziçi University, 2017\\B.S., Mathematics, Boğaziçi University, 2017\\
	M.S., Computer Science, Yale University, 2020\\
	MBA, Quantic School of Business and Technology, 2021}
	
\author{Kerim Gökarslan}
\program{Computer Engineering}
\subyear{2022}

\supervisor{Prof. Tuna Tuğcu}
\examineri{Prof. Berk Canberk}
\examinerii{Prof. Ali Emre Pusane}

\dateofapproval{03.06.2022}

\begin{document}
\pagenumbering{roman}
\makemstitle % M.S. thesis
\makeapprovalpage
\newpage
\mbox{}
\vfill
\begin{flushright}
    \par\null\par\par\null\par\par\null\par\par\null\par\par\null\par\par\null\par
    % \selectlanguage{greek}
    % \textit{ὁ δὲ ἀνεξέταστος βίος οὐ βιωτὸς ἀνθρώπ}\\
    % \selectlanguage{english}
    % \textit{(The unexamined life is not worth living for a human being.)}\\ \textit{Σωκράτης (Socrates)}\\
    % \selectlanguage{english} \par\null\par
    % \textit{Dünyada her şey için, maddiyat için, maneviyat için, hayat için, başarı için en hakikî yol gösterici ilimdir, fendir. İlim ve fennin dışında yol gösterici aramak gaflettir, cahilliktir, doğru yoldan sapmaktır.}
    % \\     \textit{(Science is the most real guide for civilisation, for life, for success in the world. To search for a guide other than science is absurdity, ignorance and heresy.)} \\
    % \textit{M. Kemal Atatürk}
    
    \selectlanguage{greek}
    \textit{``ὁ δὲ ἀνεξέταστος βίος οὐ βιωτὸς ἀνθρώπ"}\\ \textit{Σωκράτης}  \\
    \selectlanguage{english} 
    \textit{``Dünyada her şey için, maddiyat için, maneviyat için, hayat için, başarı için en hakikî yol gösterici ilimdir, fendir. İlim ve fennin dışında yol gösterici aramak gaflettir, cahilliktir, doğru yoldan sapmaktır."}
    \\
    \textit{M. Kemal Atatürk}
    \par\null\par\par\null\par\par\null\par\par\null\par
    \textit{``The unexamined life is not worth living for a human being."}\\ \textit{Socrates}\\
    
   \textit{``Science is the most real guide for civilisation, for life, for success in the world. To search for a guide other than science is absurdity, ignorance and heresy."} \\ 
    \textit{M. Kemal Atatürk}

\end{flushright}
\begin{acknowledgements}
This thesis and most of my academic achievements would not be possible if I didn't have the chance to meet with Prof. Tuna Tugcu back in 2012 when I was a senior high school student to work on a then cutting-edge Bluetooth/GPS cooperation project that we developed for the TUBITAK's high school project contest.
He led me to the way that I ended up at Yale, and thanks to him, I truly understood what I like and what I don't about science and engineering. And today, after two bachelor's degrees and being a candidate for my third master's degree, I look back on my ten years, and I feel highly empowered to achieve tomorrow. 

Further, I would like to thank Prof. Ali Emre Pusane and Prof. Arda Yurdakul for their letters of recommendation on my way to Yale. I thank all of the professors I work with, even though numerous of them only taught me what professorship is not about and how professors should not act.

And my personal life: I want to thank all of my friends, from high school to Bogazici, then at Yale, and finally at the companies I worked for. Most importantly, our extraordinary, mad, and yet extremely successful group, \textit{İSTİKRAR}. Of course, I would not be the person today if it weren't for my parents and brother. They have always had my back and supported me no matter what, and I am glad to have their sincere support.
\end{acknowledgements}
\begin{abstract}
Recent advancements in wireless technologies towards the next-generation cellular networks have brought a new era that made it possible to apply cellular technology on traditionally-wired networks with tighter requirements, such as industrial networks. The next-generation cellular technologies (e.g., 5G and Beyond) introduce the concept of ultra-reliable low-latency communications (URLLC). This thesis presents a Software-Defined Networking (SDN) architecture with programmable data planes for the next-generation cellular networks to achieve URLLC. Our design deploys programmable switches between the cellular core and Radio Access Networks (RAN) to monitor and modify data traffic at the line speed. We introduce the concept of \textit{intra-cellular optimization}, a relaxation in cellular networks to allow pre-authorized in-network devices to communicate without being required to signal the core network. We also present a control structure, Unified Control Plane (UCP), containing a novel Ethernet Layer control protocol and an adapted version of link-state routing information distribution among the programmable switches. Our implementation uses P4 with an 5G implementation (Open5Gs) and a UE/RAN simulator. We implement a Python simulator to evaluate the performance of our system on multi-switch topologies by simulating the switch behavior. Our evaluation indicates latency reduction up to 2x with \textit{intra-cellular optimization} compared to the conventional architecture. We show that our design has a ten-millisecond level of control latency, and achieves fine-grained network security and monitoring.
\end{abstract}
\begin{ozet}

Yeni nesil kablosuz hücresel ağlardaki son gelişmeler, hücresel teknolojinin endüstriyel ağlar gibi daha ağır gereksinimleri olan ve geleneksel olarak kablolu ağlar kullanan sistemlerde uygulanmasını mümkün kılan yeni bir dönem getirdi. Yeni nesil hücresel teknolojiler (ör. 5G ve Ötesi), aşırı güvenilir düşük gecikmeli iletişim (URLLC) konseptini ortaya çıkardı. Bu tezde, URLLC'yi elde etmek için yeni nesil hücresel ağlarda programlanabilir veri düzlemleriyle Yazılım Tanımlı Ağ (SDN) mimarisini öneriyoruz. Tasarımımız,  veri trafiğini hat hızında kontrol etmek için hücresel çekirdek ile Radyo Erişim Ağı (RAN) arasında programlanabilir ağ anahtarları kullanıyor. Önceden yetkilendirilmiş ağ içi cihazların çekirdek ağa sinyal göndermesi gerekmeden iletişim kurmasını sağlamak için bir gevşetme olan \textit{hücreler arası eniyileme (eng. intra-cellular optimization)} kavramını tanıtıyoruz. Ayrıca, programlanabilir anahtarlar arasında bilgi dağıtımı için tasarladığımız Birleştirilmiş Kontrol Düzlemi (eng. Unified Control Plane (UCP)) protokolünü sunuyoruz. Mimarimizi Open5Gs uygulaması, bir UE/RAN simülatörü ve P4 programlama dili kullanarak uygularken sistemimizin çoklu anahtar topolojilerindeki performansını geliştirdiğimiz Python simülatörü ile değerlendiriyoruz. Kapsamlı değerlendirmemiz, geleneksel mimariye kıyasla \textit{hücreler arası eniyileme} ile gecikme süresinin iki kata kadar azaldığını gösteriyor.
\end{ozet}
\tableofcontents
\listoffigures
\listoftables
\begin{symbols}
% The title will be typeset as "LIST OF SYMBOLS".
%
% Use a separate \sym command for each symbols definition.
% First, Latin symbols in alphabetical order

\sym{$gnb_{ij}$}{gNB with identifier $i$}
\sym{$l_{ij}$}{A link between nodes $n_i$ and $n_j$}
\sym{$N$}{Network}
\sym{$n_{i}$}{Node with identifier $i$}
\sym{$P_{ij}$}{A path between nodes $n_i$ and $n_j$}
\sym{$\mathbb{Q}^{\geq 0}$}{The set of non-negative rational numbers}
\sym{$sw_{ij}$}{Switch with identifier $i$}
\sym{$t_{i}$}{TEID with identifier $i$}
\sym{$u$}{UPF}
\sym{$ue_{ij}$}{UE with identifier $i$}
%\sym{$\mathbf{A}$}{State transition matrix of a hidden Markov model}
% 1 EMPTY LINE BETWEEN LATIN AND GREEK SYMBOLS GROUP!!!
% Then Greek symbols in alphabetical order
%\sym{$\alpha$}{Blending parameter \textit{or} scale}
%\sym{$\beta_t(i)$}{Backward variable}
%\sym{$\Theta$}{Parameter set}
%\sym{ }{}

\end{symbols}

\begin{abbreviations}
 % Abbreviations in alphabetical order
 
%\sym{}{}  
\sym{1G}{First Generation Cellular Networks}
\sym{2G}{Second Generation Cellular Networks}
\sym{3G}{Third Generation Cellular Networks}
\sym{3GPP}{3rd Generation Partnership Project}
\sym{4G}{Fourth Generation Cellular Networks}
\sym{5G}{Fifth Generation Cellular Networks}
\sym{6G}{Sixth Generation Cellular Networks}
\sym{AF}{Application Function}
\sym{AI}{Artificial Intelligence}
\sym{AMF}{Access and Mobility Management}
\sym{AN}{Access Network}
\sym{AUSF}{Authentication Server Function}
\sym{BMV2}{Behavioral Model version 2}
\sym{CAPEX}{Capital Expenditures}
\sym{CDR}{Charging Data Record}
\sym{CHF}{Charging Function}
\sym{CMI}{Control Message Identifier}
\sym{CN}{Core Network}
\sym{CPU}{Central Process Unit}
\sym{CUPS}{Control and User Planee Seperation}
\sym{DN}{Data Network}
\sym{ETSI}{European Telecommunications Standard Institute}
\sym{FSM}{Finite State Machine}
\sym{GHz}{Gigahertz}
\sym{GPRS}{General Packet Radio Service}
\sym{GTP}{GRPS Tunneling Protocol}
\sym{Gbps}{Gigabits per second}
\sym{HTTP}{Hypertext Transport Protocol}
\sym{ICMP}{Internet Control Message Protocol}
\sym{ID}{Identity Document}
\sym{IIoT}{Industrial Internet of Things}
\sym{IP}{Internet Protocol}
\sym{IPv4}{Internet Protocol version 4}
\sym{IPv6}{Internet Protocol version 6}
\sym{L2}{Layer 2 (Data Link Layer)}
\sym{L3}{Layer 3 (Network Layer)}
\sym{L4}{Layer 4 (Transport Layer)}
\sym{LPM}{Longest Prefix Match}
\sym{LSR}{Link State Routing}
\sym{LTE}{Long Term Evolution}
\sym{M2M}{Machine-to-Machine}
\sym{MAC}{Medium Access Control}
\sym{MME}{Mobility Management Entity}
\sym{MTU}{Maximum Transfer Unit}
\sym{Mbps}{Megatbits per second}
\sym{NEF}{Network Exposure Function}
\sym{NF}{Network Function}
\sym{NG}{Next Generation}
\sym{NGAP}{NG Application Protocol}
\sym{NR}{New Radio}
\sym{NRF}{Network Repository Function}
\sym{NSSF}{Network Slice Selection Function}
\sym{NWDAF}{Network Data Analytics Function}
\sym{OPEX}{Operating Expenses}
\sym{OS}{Operating System}
\sym{OSPF}{Open Shortest Path First}
\sym{PCF}{Policy Control Function}
\sym{PCRF}{Policy and Charging Rules Function}
\sym{PDU}{Protocol Data Unit}
\sym{QoS}{Quality of Service}
\sym{RAM}{Random Access Memory}
\sym{RAN}{Radio Access Network}
\sym{RTT}{Round-trip Time}
\sym{SCP}{Serrvice Communication Proxy}
\sym{SCTP}{Stream Control Transmission Protocol}
\sym{SDN}{Sfotware-Defined Networking}
\sym{SMF}{Session Management Function}
\sym{SW}{Switch}
\sym{TCAM}{Ternary Content-Addressable Memory}
\sym{TCP}{Transmission Control Protocol}
\sym{TEID}{Tunnel Endpoint Identifier}
\sym{UCP}{Unified Control Plane}
\sym{UDM}{Unified Data Management}
\sym{UDP}{User Datagram Protocol}
\sym{UDR}{Unified Data Repository}
\sym{UE}{User Equipment}
\sym{UPF}{User Plane Function}
\sym{URLLC}{Ultra-Reliable and Low-Latency Communication}
\sym{VM}{Virtual Machine}
\sym{VR}{Virtual Reality}
\sym{WiMAX}{Nobile Worldwide Interoperability for Microwave Access}
\sym{dst}{Destination}
\sym{eMBB}{Enhanced Mobile Broadband}
\sym{gNB}{Next Generation NodeB}
\sym{mMTC}{Massive Machine Type Communications}
\sym{ms}{Milliseconds}
\sym{s}{Seconds}
\sym{src}{Source}
\sym{vCPU}{Virtual Central Process Unit}

% \sym{\tablefootnote{In this thesis, we assume all access networks are radio access networks; thus, we simply refer to them as RAN unless otherwise is noted.} 
\end{abbreviations}

\chapter{INTRODUCTION} \label{chapter:introduction}
\pagenumbering{arabic}

Wireless cellular systems have been rapidly evolving since the first generation (1G) based on analog voice communication was introduced in the late 1970s. The first digitalized cellular technology, the second-generation (2G), was introduced a decade later. The following extensions of 2G (e.g., GPRS) initially enabled the cellular networks to reach data rates around a couple of tens of Kbps in measurements~\cite{Korhonen2001}, and later with introduction of EDGE such networks can reach upto 384 Kbps data rates. In the late 2000s, the third-generation (3G) cellular technology with transmission rates reaching Mbps was introduced. The 3G technology supports direct connection to Transmission Control Protocol/Internet Protocol (TCP/IP) based services, including the Internet~\cite{Tan2008}. As this was the turning point for cellular networks to transfer data rather than voice or video, the pressure to have better latency and data rate values increased. Consequently, the 3rd Generation Partnership Project (3GPP) has presented the standardization of the fourth-generation (4G) cellular systems, including the Long Term Evolution (LTE) along with Mobile Worldwide Interoperability for Microwave Access (WiMAX), which can offer data rates up to 50 Mbps~\cite{Krapichler2007}. Today, we see 5G deployments worldwide aiming to reach Gbps data rates with milliseconds-level latency values.

As recent developments in cellular technologies are the first steps for numerous latency-sensitive applications to switch to wireless networks, the future cellular technology generations (i.e., 5G and Beyond) have much room to improve latency while ensuring reliability. We, in fact, consider this trend within 5G to bring newer radio and core network design that aims at several use cases, including ultra-reliable low-latency communications (URLLC) aiming at mission-critical services such as industrial networks. With the introduction of such systems to cellular networks, we have started to see the trend of replacing wired networks with a cellular equivalent that targets to achieve similar requirements. Switching to wireless systems is indeed more profitable for network providers as wireless technologies can reduce both operating expenses (OPEX) (e.g., reduced maintenance costs) and capital expenditures (CAPEX) (e.g., planning and infrastructure costs) in many wired-network systems. Nevertheless, today's cellular network technology is far from the expectation of 99.9999\% reliability with 1 ms latency~\cite{Li2018}.

Initially introduced as a part of 5G New Radio (NR) by the 3GPP in the first complete set of 5G standards (Release 15)~\cite{ETSI123501v152}, URLLC aims to achieve 1 ms latency and target latency-sensitive applications, including industrial networks, smart grids, tactile internet~\cite{tactile}, and autonomous driving. URLLC thus promises to be one of the critical features of not only the 5G but also the next generations of cellular networks. Unfortunately, achieving URLLC is not straightforward because of both the physical layer requirements and cellular networks' current design. Apart from physical layer drawbacks, the current cellular network designs have numerous latency-unaware way of working that causes additional delays. This thesis focuses on low-latency applications at the packet processing level. It introduces a programmable data plane pipeline for URLLC in cellular networks to achieve lower latency. It targets 5G and Beyond cellular networks by introducing the concept of pre-authorization where some of the cellular network functions (e.g., Access and Mobility Management Function (AMF), Session Management Function (SMF)) are offloaded. We do so by leveraging the recent advancements in Software-Defined Networking (SDN), specifically using programmable data planes. In recent years, programmable data planes proved their value mainly in a data center or service-provider networks. In this thesis, we take a new approach to offload latency in cellular networks with higher intra-cellular communication (e.g., industrial cellular networks) using programmable data planes.

The principal idea of SDN is the decoupling of data and control paths into different devices to achieve more flexible and robust networks. In many SDN systems, the control plane runs as a centralized system that manages the network equipment responsible for the data plane in real-time. Further, these data plane devices have limited processing capabilities, such as processing pre-defined types of packets. In recent years, researchers have introduced the concept of programmable data planes, where the data planes devices can be programmed so that they can have customized behavior with the ability to process packets at the line rate. This consequently allows network administrators to define application-aware network behavior such as achieving latency-guarantees~\cite{Bifulco2018}. In this thesis, we use the P4 language~\cite{bosshart2014p4} to implement our cellular-specific data plane (i.e., our data plane can parse and process cellular protocols such as GPRS Tunneling Protocol (GTP)) to achieve the aforementioned cellular network design and demonstrate it on a software switch. 

\begin{figure}[htbp]
    \centering
    \includegraphics[width=0.9\columnwidth,keepaspectratio]{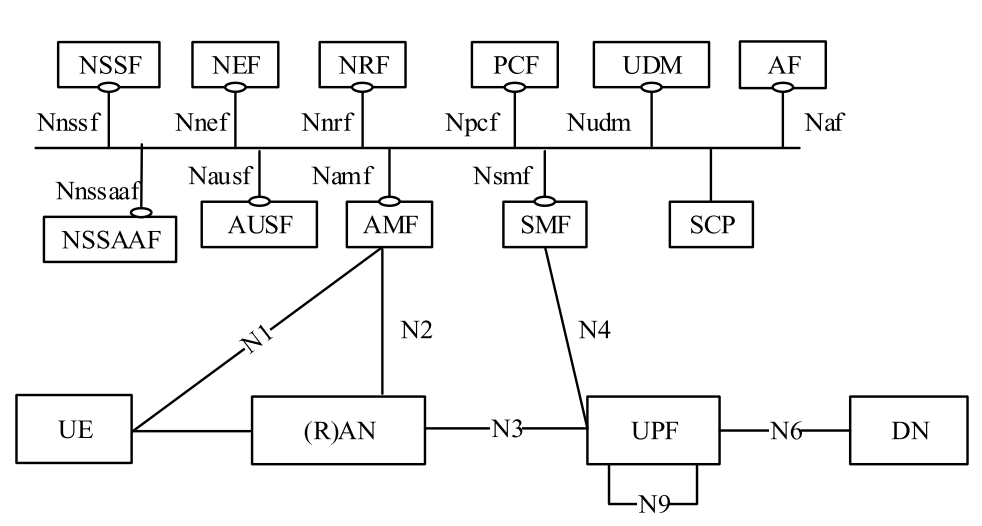}
    \caption{A simplified version of the 5G Network Architecture from the ETSI Standard Number TS 123 501 Release 16~\cite{ETSI2020a}.}
    \label{fig:5g-arch}
\end{figure}

\section{Contributions of This Thesis}
This thesis aims to achieve URLLC on 5G and Beyond cellular networks. While our high-level design can be applied to the newer generations of cellular networks such as 5G, we use the 5G - Release 16 version in our implementation; therefore, we discuss our implementation within the 5G specifications. As shown in Figure~\ref{fig:5g-arch}, 5G architecture consists of three main components: (a) User Equipment (UE); (b) Radio Access Network (RAN), and 5G core functions; and (c) User Plane Function (UPF) and Data Network (DN). The UE connects to the cellular network via RAN and communicates with other devices/outside world via UPF. RAN connects (via N2) to AMF, and UPF connects (via N4) to SMF on the cellular control path that is responsible for signaling including a UE initialization or handover. The traffic from and to the UE uses GTP over IP while it is carried between UE, RAN, and UPF. As we mainly focus on the cellular data path, this thesis does not discuss the other 5G core functions in detail.

The contributions of this thesis, therefore, are:

\begin{itemize}
    \item Introduction of the concept of intra-cellular optimization with cellular pre\-au\-thor\-i\-za\-tion to significantly reduce the latency between two cellular devices in the same cellular network, 
    \item Achieving data plane-assisted network security without requiring any modifications on the cellular network hardware or software,
    \item Introducing fine-grained network monitoring for cellular networks via pro\-gram\-ma\-ble data planes,
    \item Introduction of a novel L2-layer protocol, Unified Control Plane (UCP), to con\-trol the programmable data plane that connects cellular network components in real-time,
    \item Introduction of a mathematical model to study intra-cellular latency behavior in cellular networks,
    \item One of the first research work that uses programmable data planes to parse both GTP and NGAP protocols.
\end{itemize}

Our evaluations and the analysis of the mathematical model of latency we develop indicate that our design can reduce data latency by a factor of two compared to traditional cellular network setups in 5G and Beyond. This optimization is practical, especially for non-traditional cellular systems such as industrial networks, where most devices (e.g., sensors) communicate within the network. One of the concerns for URLLC is the possibility of reduced network security and traceability; we show that our design can achieve detailed network monitoring and security using the capabilities of programmable data planes.

The rest of this thesis is organized as follows. Chapter~\ref{chapter:relatedwork} discusses the overview of the related work on cellular networks, SDN, and URLLC. Chapter~\ref{chapter:architecture} presents our novel architecture, including the UCP and the programmable data path. Chapter~\ref{chapter:implementation}, then, discusses the details of our implementation. Chapter~\ref{chapter:evaluation} gives the evaluation results. Chapter~\ref{chapter:conclusion}, finally, concludes the thesis with the future research directions.

\chapter{RELATED WORK}\label{chapter:relatedwork}

\section{Cellular Networks: 5G and Beyond}

Researchers categorize 5G and Beyond cellular networks under three categories: massive machine type communication (mMTC) addresses the applications with high data rates, enhanced mobile broadband (eMBB) concentrates on machine-to-machine (M2M) communication, and URLLC addresses low latency and highly available services including industry-grade cellular networks and tactile internet~\cite{popovski20185g, tactile}.

As many countries have already started to deploy 5G networks, researchers have been working on the next-generation cellular networks, namely 5G and Beyond as well as the sixth generation (6G). A survey on 6G networks indicates that such next-generation networks are needed to achieve more cost-efficient and better quality of the use cases introduced in 5G, such as IoT, virtual reality (VR), and autonomous driving~\cite{jiang2021road}. A recent survey~\cite{habibi2019comprehensive} on RAN architectures in next-generation networks indicates SDN as a key to realizing high efficiency and high flexibility, and consequently, we focus on achieving URLLC with SDN in our work.

\section{URLLC}

As introduced with the 5G NR in the 3GPP standards URLLC targets non-traditional cellular networks aiming latency values on the order of milliseconds such as tactile internet, autonomous driving, and industrial networks. Numerous studies focus on the scheduling aspect of cellular networks to achieve URLLC. The authors of~\cite{Anand2020} study a joint scheduling schema for URLLC and eMBB aiming broadband utility while having URLLC requirements. Claiming to be the first formalization of the joint URLLC and eMBB scheduling, they propose a scheduling framework with linear, convex, and threshold models with theoretical guarantees on the algorithms. 

Li et al. take a different approach on the same problem and use deep reinforcement learning to schedule URLLC and eMBB~\cite{Li2020}. They use the deep deterministic policy gradient method to learn the tradeoff policy regarding reliability and service satisfaction. 

\cite{y3} discusses URLLC techniques, including the AI-enabled edge caching frameworks focused on caching. Further, it discusses one of the most popular URLLC techniques, grant-free access, that decreases the control overhead via dynamic allocation of the resources. The authors of \cite{y3} also present a comprehensive study of URLLC while lacking SDN applications to achieve URLLC. Researchers in \cite{y4} examines the URLLC on 5G Releases 15, 16, and 17. The authors of ~\cite{y2} discuss diversity reception, automatic repeat query, and cognitive radio algorithm for 5G URLLC at a 28GHz link to overcome link degradation and interference to achieve lower latencies. 

There are also a few studies discussing real-life use cases of URLLC technology. One such study in~\cite{Yogapratama2020} discusses the case of an Indonesian Operator. They use the measurements of the existing network to propose the path to achieve the URLLC target of 1 ms with a design of data center locations. Another study discusses a scheme for factory automation targetting the resource allocation problem that enables autonomous driving within the factory~\cite{Jayaweera2020}.

\section{SDN and Programmable Data Planes}

SDN brings the concept of data plane and control plane separation, where the former contains the user data traffic and the latter contains the network control signals. Researchers have extensively applied SDN techniques to wireless networks in the last decade. SDN brings the concept of data plane and control plane separation, where the first one contains the user data traffic and the second one contains the network control signals. This separation, therefore, allows deployment of control plane to be centralized and the use of distributed data plane over the network. 

SDN increases both interoperability and manageability of devices, as it allows generic data plane switches (e.g., OpenFlow~\cite{openflow}). When introducing a new protocol, such devices do not need a new hardware design, as they can be defined via software. This consequently decreases the development time and cost significantly.

One of the classical SDN design limitations is the limit of the data plane behavior. Many existing solutions focus on the operation of the 5-tuple of TCP/IP, thus lacking packet processing capabilities at the application layer. Programmable data planes have materialized to define user-specific data plane behavior at the line rate. Programmers can define customizable data paths with specific programmable data plane languages. One of such languages, P4~\cite{bosshart2014p4}, has been extensively used in academia and industry in the last decade~\cite{kfoury2021exhaustive, Bifulco2018}. This thesis uses P4 language to implement a software switch's cellular network-specific data plane.

\section{SDN in Cellular Networks}

Researchers have extensively applied the SDN techniques to wireless networks in the last decade~\cite{Sun2015, Prados-Garzon2016, Ricart-Sanchez2019}. In~\cite{Zaidi2018}, researchers indicate that SDN plays a crucial role in achieving a flexible and virtualized 5G network, although the many existing solutions are not SDN-ready, i.e., most SDN applications in the cellular networks are work-in-progress. The authors of~\cite{y11} and~\cite{y12} study network slicing on a 5G core network via decoupling UPF and SMF.~\cite{y12} considers the dynamic traffic and structures the UPF programmable module integrated with SDN to give traffic flow rules to achieve network slicing in URLLC. It also introduces a different programmable switch to control the network flow and assure network security instead of replacing network functions with other programmable switches such as UPF or SMF. The authors of~\cite{y11} do not consider any performance criteria such as latency, while~\cite{y12} aims to decrease the latency with security considerations on private industrial 5G  networks.  

\newpage 

Many studies\cite{y5,y6} focuses to the user behavior and UPF with SDN on mobile environments. Researchers predict the user behavior via an SDN containing a foreseeing functionality of the UPF placement to reduce the user plane configuration delay. They further propose pre-configured UPF as a handover mechanism~\cite{y5} uses the intermediate UPFs to anticipate the user behavior on SMF with a learning-based approach.

\chapter{ARCHITECTURE}\label{chapter:architecture}

In this chapter, we introduce the cellular network architecture with programmable switches. The first section describes the state-of-the-art 5G architecture and related network functions that we take as a basis for our implementation. Considering a minimal-to-none change in the cellular network architecture and implementation, we introduce our system running as a single P4-program on the P4 switches interconnecting the gNBs and core NFs as shown in Figure~\ref{fig:5g-arch} in Section~\ref{sec:arch-highlevel}. The key idea for our design is to achieve URLLC while being seamless to users of the cellular network, i.e., users do not change their cellular network hardware or software. After a high-level introduction of our architecture in Section~\ref{sec:arch-highlevel}, we investigate our design in detail with three parts: (a) the unified control plane (UCP) that controls the P4 switches real-time using the high-level configuration from the network administration (Section~\ref{sec:arch-ucp}). (b) the cellular control plane, where the P4 switch processes the control messages between gNB and AMF using NGAP (Section~\ref{sec:arch-cp}), (c) the cellular data plane, where the P4 switch processes the data packets between gNBs and UPF encapsulated using GTP (Section~\ref{sec:arch-dp}). We finally introduce our mathematical model for intra-cecullar latency estimation in cellular networks, which we use in our evaluation results in Section~\ref{sec:arch-math}.

\section{Cellular Network Architecture and Functions}\label{sec:arch-5g}

In this section, we discuss the details of the 5G cellular network architecture since we take 5G architecture as the basis in our implementation. However, our architecture design is independent of the  5G network and can be adapted into newer cellular network generations. The Mobility Management Entity (MME) in 4G architecture, which is responsible for session and mobility management as well as the UE authentication, is split into three functions in 5G, AMF, SMF, and Authentication Server Function (AUSF)~\cite{Shepherd2018}. As our programmable data plane processes the signaling between gNBs and these control functions, we investigate the characteristics of the AMF, SMF, and AUSF, and partially implementing their functionality in the data path as described in Section~\ref{sec:arch-cp}. We further discuss the potential extensions in Chapter~\ref{chapter:conclusion}. The final subsection studies the cellular network's data plane function, i.e., the UPF. Our programmable data path does not implement UPF itself; it introduces intra-cellular network optimization. The communication between two users in the same 5G network is processed at the switch interconnecting respective gNBs rather than the UPF to reduce latency. 

\makeatletter
\setlength{\@fptop}{0pt}
\makeatother
\begin{figure}[htbp]
    \centering
    \includegraphics[width=\columnwidth,keepaspectratio]{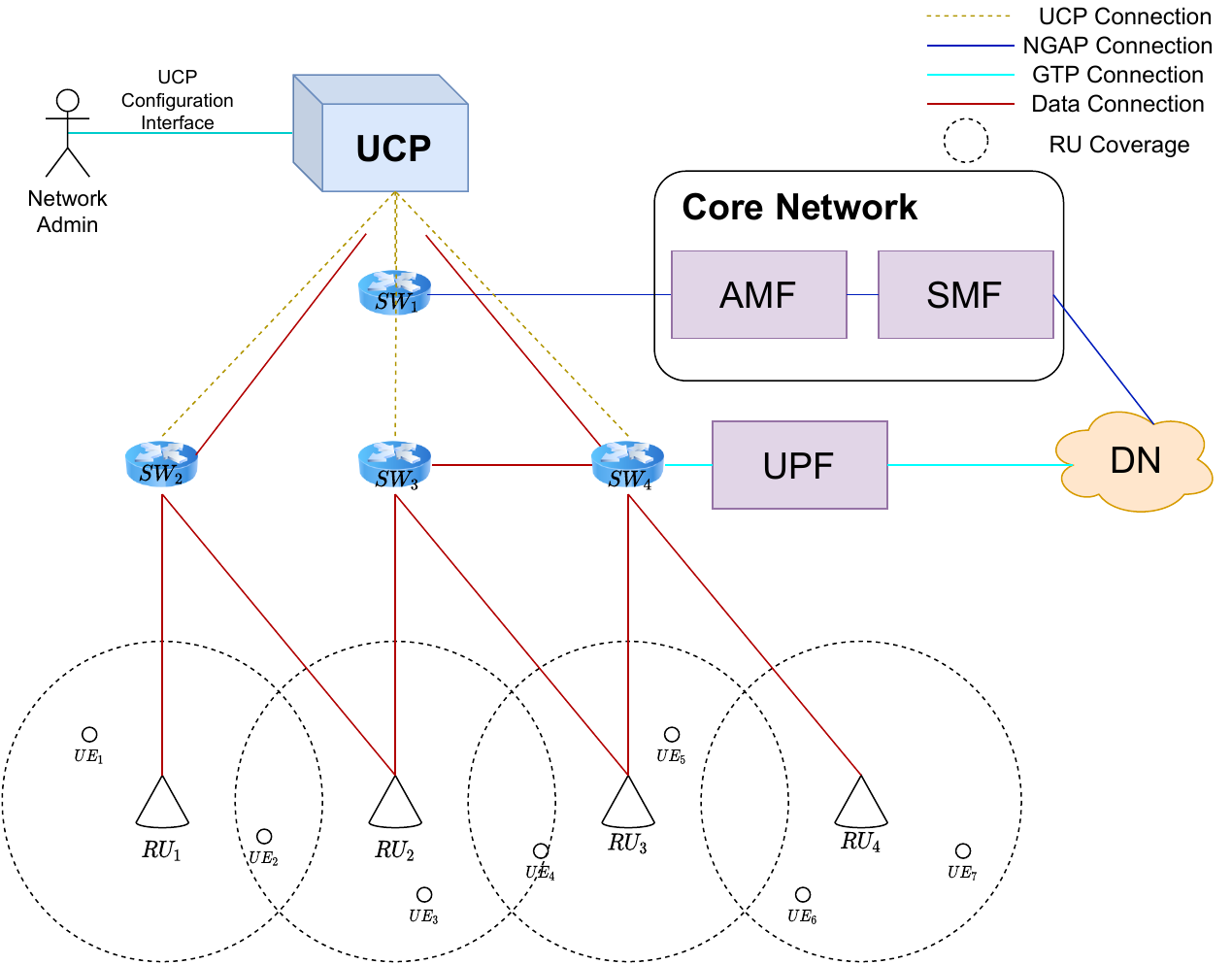}
    \caption{An architecture with four gNBs ($gNB_i$) with the same signal coverage (dashed-lined circles). Four programmable switches are connecting them to the cellular network, and seven devices ($UE_i$). Network administrators manage switches via the UCP logically connected to the P4 switches in real-time.}
    \label{fig:arch-detail}
\end{figure}

\newpage 

\subsection{Cellular Control Plane}

As our programmable data plane design offloads some of the UE and cellular CN signaling, we describe related 5G NFs (i.e., AMF and SMF) and the NG Application Protocol (NGAP), which is the protocol for control signaling between 5G RAN and CN, in this section. We, furthermore, discuss other potential behaviors and NFs that can be offloaded by programmable design in Chapter~\ref{chapter:conclusion}.

\subsubsection{NGAP}

Specified by 3GPP, the NGAP defines the control signaling between RAN elements (i.e., UE and gNB) and cellular core via AMF~\cite{ETSINGAP}. NGAP consists of a wide variety of procedures, from PDU session establishment~\cite{ETSI138415v161} and handover control to cell resource coordination. In this thesis, we focus on NGAP signaling for individual UE and process UE initialization packets. As our work mainly focuses on the cellular data plane, the NGAP processing only registers the IDs of UE with its respective gNB.

\subsubsection{AMF}

AMF is mainly responsible for handling mobility and connection management of the UE while it is the single point of contact with 5G-RAN, and consequently, the UE. AMF forwards the information related to UE to session management SMF using the N11 link. 

\subsubsection{SMF}

\begin{figure}[htbp]
    \centering
    \includegraphics[width=\columnwidth,keepaspectratio]{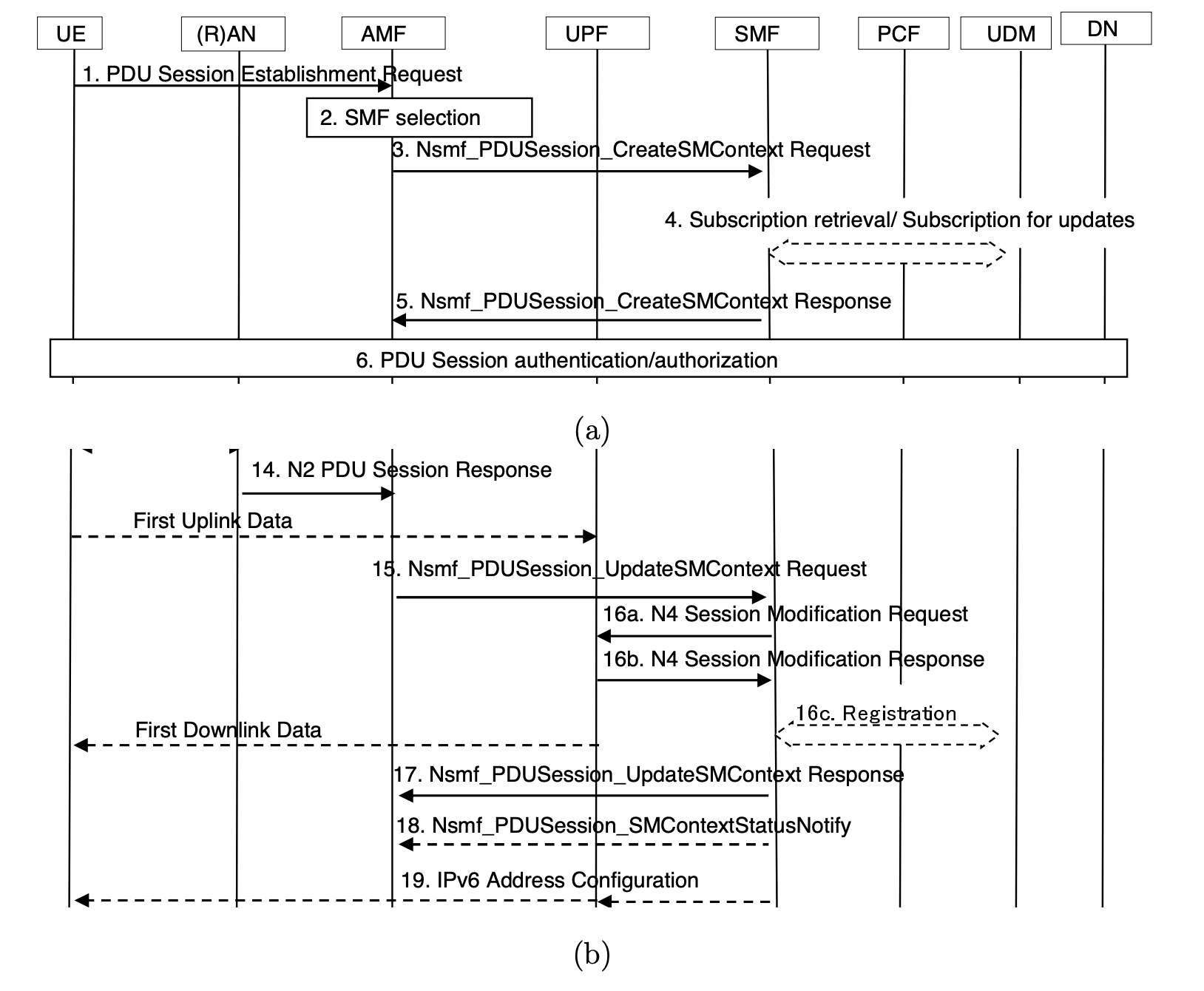}
    \caption{A simplified version (steps between 5-14 and 10-21 are not shown) of PDU session establishment initiated by the UE  in ETSI TS 123 502 and ETSI 138 300 Release 16~\cite{ETSI123502v165, ETSI138300v162}.}
    \label{fig:pdu-establishment}
\end{figure}

The SMF's principal responsibility is to manage PDU sessions with the UPF via the N4 interface introduced with 5G. The UE connects to the cellular DN via UPF once the PDU session is established. A simplified UE initialization signalling for PDU session establishment is shown in Figure~\ref{fig:pdu-establishment}.

\makeatletter
\setlength{\@fptop}{0pt}
\makeatother

\subsection{Cellular Data Plane} 

The UE data packets are encapsulated with GTP and transmitted between gNBs and UPF using N3, N6, and N9 interfaces as shown in Figure~\ref{fig:5g-arch}.  

\subsubsection{GTP} 

Unlike the aforementioned protocols and network functions, GTP has been used since the introduction of GPRS\cite{3gpp.29.060}.
GTP is a protocol run on top of the TCP/IP or UDP/IP, and used to carry the UE data between RAN and core networks. GTP has two major versions, where older cellular generations deploy GTPv1 and newer generations, including 5G, use GTPv2. In addition to encapsulating user data, GTP is also used to check connectivity via echo messages. 

\begin{table}[thbp]
\vspace{1em}
\begin{center}
\caption{GTPv2 header fields defined in the 3GPP standard~\cite{3gpp.29.274}.} 
\vspace{0.5em}
\begin{ttfamily}
\small
\begin{tabular}{|c|c|c|}
    \hline
    First Bit & Length (bit) & Field Description \\ \hline
    0 & 3 & Version \\ \hline
    3 & 1 & Piggybacking Flag \\ \hline
    4 & 1 & TEID Flag ($T$) \\ \hline
    5 & 3 & Spare \\ \hline
    8 & 8 & Message Type \\ \hline
    16 & 16  & Message Length  \\ \hline
    32 & 32 & TEID when $T = 1$ \\ \hline
    $\begin{cases}
      64, & \text{if}\ T = 1 \\
      32, & \text{otherwise}
    \end{cases}$ & 24 & Sequence Number \\ \hline
\end{tabular}
\end{ttfamily}
\label{tab:gtp-header}
\end{center}
\end{table}

As shown in Table~\ref{tab:gtp-header}, the GTP header contains the information about the user data that is encapsulates, and also includes the Tunnel Endpoint Identifier (TEID) that describes the tunnel between UE and gNB. Except for the echo messages, all GTP packets belong to a UE, and therefore GTP header carries the associated TEID. The sequence number indicates the order of the packet within the respective tunnel.

\subsubsection{UPF}
The 5G architecture brings the concept of control and user plane separation (CUPS)~\cite{3gpp.23.214} to decouple the control functions from the data plane functions so that the network can achieve higher data rates and more complex controls, including network slicing. The UPF is the function responsible for user-plane communication with the DN. UPF is responsible for encapsulation and decapsulation of the GTP-encapsulated UE data, and the UE communicates outside of the cellular network via the UPF. The UPF also communicates with SMF via the N4 interface, where SMF reports the PDU sessions. UPF furthermore deploys the flow-grained quality of service (QoS) rules. 

\section{High-Level View}\label{sec:arch-highlevel}

\begin{figure*}[!ht]
    \vspace{2em}
    \centering
    \includegraphics[width=\linewidth]{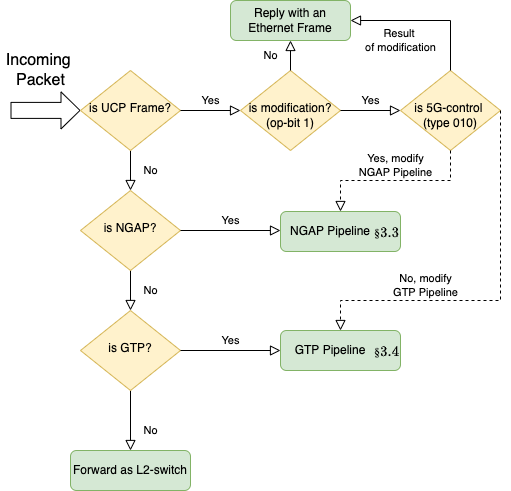}
    \caption{The process flow of the programmable data plane with the RAN and core network as appeared in our work~\cite{Gokarslan2022}.}
    \label{fig:packet-flowchart}
\end{figure*}

\newpage 
This section discusses the high-level view of our programmable data path and its control protocol, UCP. As shown in Figure~\ref{fig:packet-flowchart}, we handle three types of packets: (a) UCP, (b) NGAP, i.e., cellular control plane, and (c) GTP, i.e., cellular data plane. We define the UCP so that network administrators can modify the rules such as network security and get network data such as user-specified traffic monitoring. We further use UCP for switch-to-switch communication.

Figure~\ref{fig:packet-flowchart}  shows the packet-level processing at the programmable switches between RAN and the cellular core network. The data plane first checks the type of incoming packet to decide further processing. If it is a UCP frame (Ethernet Type $\mathbf{0xf1f1}$), the data plane checks its 8-bit CMI to understand the type of UCP frame. Table~\ref{tab:p4control-operations} shows the supported types of UCP frames, including their description and payload. The data plane finally replies with a UCP-reply frame if the received UCP frame requires a reply. 

Suppose there is an incoming packet to the switch.  If the switch identifies this packet as an NGAP packet (NGAP uses IP over SCTP with source or destination port $\mathbf{38412}$), it processes this packet using the NGAP pipeline described in Section~\ref{sec:arch-cp}.  If it identifies the packet as a GTP packet (GTP uses IP over UDP with source port $\mathbf{2152}$), then the switch processes  it using the GTP pipeline described in Section~\ref{sec:arch-dp}.  Finally,  if the packet does not match any of the aforementioned types,  the switch behaves as an L2-switch,  i.e.,  forwards the packet using the MAC table.

\subsection{Unified Control Plane (UCP)}\label{sec:arch-ucp}

The UCP uses a layer-2 protocol that we define to manage the P4 switches between RAN and the core network. The control message is sent as Ethernet frames having an 8-bit control message identifier (CMI). The first 3 bits of CMI describe the operation type, and the following 5 bits describe the operation. The rest of the frame carries operation-related data. We implement the operations shown in Table~\ref{tab:p4control-operations}.

\subsubsection{Multi-Switch Communication} 

Many cellular networks contain hundreds or thousands gNBs, requiring multi-switch topology to connect them to the core networks. 

\begin{table}[t!]
\centering
\caption[UCP Message Types]{UCP messages are Ethernet Frames starting with an 8-bit CMI (first 3 bits for operation type and last 5 bits for operation ID) followed by operation data if required. P4 switch generates a reply message and sends it back to the requester with a descriptor indicating the status. Note that 8-bit switch ID is always added to the frame data for the frames from and to a switch.} \label{tab:p4control-operations}
\vspace{0.5em}
%\footnote{Ethernet Frame, IPv4, TCP, UDP, and ICMP is supported. For example, if a user wants to get monitoring stats for SSH packets to a specific MAC address, it should append a TCP packet header with destination port number and destination MAC address. Otherfields should be 0}

\begin{tabular}{|c|c|c|}
\hline
Operation Type & Op. ID & Description\\ \hline
\multirow{4}{*}{000 - Security} & {00000} & {Get all whitelisted IPv4s} \\ \cline{2-3}
& {00001} & {Get all blacklisted IPv4s} \\ \cline{2-3}
& {10000} & \makecell{Add an IPv4 to whitelist} \\ \cline{2-3}
& {10001} & \makecell{Add an IPv4 to blacklist} \\ \hline

\multirow{4}{*}{001 - Monitoring} & {00000} & \makecell{Get monitoring stats} \\ \cline{2-3}
& {00001} & \makecell{Get monitoring rule} \\ \cline{2-3}
& {00010} & {Get number of monitoring rules} \\ \cline{2-3}
& {10000} & \makecell{Add a monitoring rule \\ (Specification as frame/packet header)} \\ \hline %\cline{2-4}

\multirow{2}{*}{010 - 5G Control} & 00000 & Get current number of UE \\ \cline{2-3}
& 10000 & \makecell{Delete UE ID}  \\ \cline{1-3}

\multirow{1}{*}{011 - 5G Data} & 10000 & \makecell{Add UE IPv4}\\ \cline{1-3}

\multirow{3}{*}{110 - gNB Control} & 10000 & \makecell{New TEID} \\ \cline{2-3}
& 10001 & \makecell{Remove TEID}\\ \cline{2-3}
& 10011 & \makecell{Path $SW_j$ to $SW_j{^\prime}$} \\ \cline{1-3}

\multirow{4}{*}{111 - Reply} & 00000 & \makecell{Reply without modification} \\ \cline{2-3}
& 10000 & \makecell{Modification succeeded} \\ \cline{2-3}
& 10001 & \makecell{Modification failed} \\ \cline{2-3}
& 10010 & \makecell{Nexthop is updated} \\ \cline{1-3}
\end{tabular}
\end{table}

In this section, we describe our interswitch protocol managing intra-cellular optimization. As we elaborate in Section~\ref{sec:arch-dp}, intra-cellular optimization allows certain UE within the same cellular network to communicate without sending packets to UPF. The packet forwarding is done within the P4 switches using the TEID of receiving UE to determine which gNB to forward. When a gNB receives a new TEID, it broadcasts the 32-bit TEID as a UCP frame with operation code $\mathbf{11010000}$ as shown in Table~\ref{tab:p4control-operations}. Our protocol uses Dijkstra's shortest path algorithm to determine the next point of TEID, and we give the pseudocode in Figure~\ref{alg:multiswitch} and Figure~\ref{alg:multiswitchucp} for a switch $SW_i$ and the UCP, respectively. UCP distributes the next-hop information in a centralized fashion by running Dijkstra's algorithm. P4 switches thus only run the TEID matching algorithm to update their TEID registers. When a path update occurs on a switch after a piece of new information from the UCP, the switch sends an acknowledgment message with code $\mathbf{11010000}$ back to the UCP. 

\begin{figure}[htbp]
\vspace{1em}
\begin{center}
\framebox[6.0in]{\begin{minipage}[t]{5.9in}
\begin{algorithmic}
\STATE On $\mathbf{SW_i}$ which has $\mathbf{n}$ number of $\mathbf{gNBs}$, $\mathbf{\{gNB_{i, j} | 1 \leq j \leq n\}}$, connected to it.
\STATE Uses the data structures $\mathbf{teids(t_k): SW_j}$ and $\mathbf{path(SW_j): \{SW_{j_1}, ...\}}$
\STATE Initially $\mathbf{teids(t_k): \emptyset}$, $\mathbf{path(SW_j): \emptyset}$

\STATE \textbf{on receive} TEID $\mathbf{t_k}$ from switch $\mathbf{SW_j}$
\begin{ALC@g}
    \STATE  $\mathbf{teids(t_k) := SW_j}$
\end{ALC@g}

\STATE \textbf{on receive} NEXTHOP $SW_{j^\prime}$ to $\mathbf{SW_j}$
\begin{ALC@g}
    \IF {$\mathbf{nexthop(SW_j)} \neq SW_{j^\prime}$}
        \STATE $\mathbf{nexthop(SW_j)} = SW_{j^\prime}$
        \STATE \textbf{unicast} $(SW_i$, \textbf{11110010}, $SW_j)$\footnote{Reply: Nexthop Updated for $SW_j$ on the switch $SW_i$.} to \textbf{UCP}
    \ENDIF
\end{ALC@g}

\STATE \textbf{on new} TEID $\mathbf{t_k}$ from $\mathbf{gNB_{i, j}}$
\begin{ALC@g}
    \STATE $\mathbf{teids(t_k) := gNB_{i, j}}$
    \STATE \textbf{broadcast} $(SW_i$,  \textbf{11010001}, $t_k)$\footnote{New TEID $t_k$ on the switch $SW_i$.} to all switches
\end{ALC@g}
\end{algorithmic}
\end{minipage}}
\end{center}
\caption[Multi-switch Intra-cellular Optimization Algorithms on a switch]{Multi-switch Intra-cellular Optimization Algorithms on switch $SW_{i}$}\label{alg:multiswitch}
% \vskip
\end{figure}

\begin{figure}[htbp]
\begin{center}
\framebox[6.0in]{\begin{minipage}[t]{5.9in}
\begin{algorithmic}
\STATE Run at \textbf{UCP}
with ordered routing path $path(SW_{j}, SW_{j^\prime})$ 
\STATE \textbf{on topology change} 
\begin{ALC@g}
    \STATE \textbf{compute} $path$ for all $(SW_{j}, SW_{j^\prime})$ pairs
    \STATE \textbf{mark} $updated(SW_{j}, SW_{j^\prime})$ if $path(SW_{j}, SW_{j^\prime})$ is changed
    %$\mathbf{\{SW_{j_1}, ..., SW_j\}}$ TEID $\mathbf{t_k}$ f%rom $\mathbf{gNB_{i, j}}$$\mathbf{|$ $\leq$  $\|\mathbf{path(SW_j)}\|$ 
    \FOR{$(SW_{j}, SW_{j^\prime})$}
    \IF{$updated(SW_{j}, SW_{j^\prime})$}
        \STATE \textbf{unicast} ($SW_j$, \textbf{11010011}, $\{SW_{j_1}, ..., SW_{j^\prime}\}$) to $SW_j$
    \ENDIF
    \ENDFOR
    %\STATE \textbf{broadcast} $(SW_i, 11010000,t_k)$\footnote{New TEID $t_k$ on the switch $SW_i$.} to all switches
\end{ALC@g}
\STATE
\STATE \textbf{on receive} REPLY message $m$
\begin{ALC@g}
    \STATE \textbf{store} $m$
\end{ALC@g}
\end{algorithmic}
\end{minipage}}
\end{center}
\caption[Multi-switch Intra-cellular Optimization Algorithms on the UCP]{Multi-switch Intra-cellular Optimization Algorithms on the UCP.}\label{alg:multiswitchucp}
\end{figure}

\section{The Cellular Control Plane}\label{sec:arch-cp} 

In addition to the cellular data plane packet processing, our architecture contains a cellular control plane packet processing within the same switch. 

Our cellular control plane program is rather a proof-of-concept work that processes NGAP packets at the line rate. It filters the PDU establishment message between the UE and RAN and stores the UE IDs on the programmable switch. We implement a type of UCP message (with message code $010xxxxx$ as seen in Table~\ref{tab:p4control-operations}) to query the live UE data from the programmable data path.
\begin{figure*}[!ht]
    \centering
    
    %columnwidth,keepaspectratio
    \includegraphics[width=\linewidth]{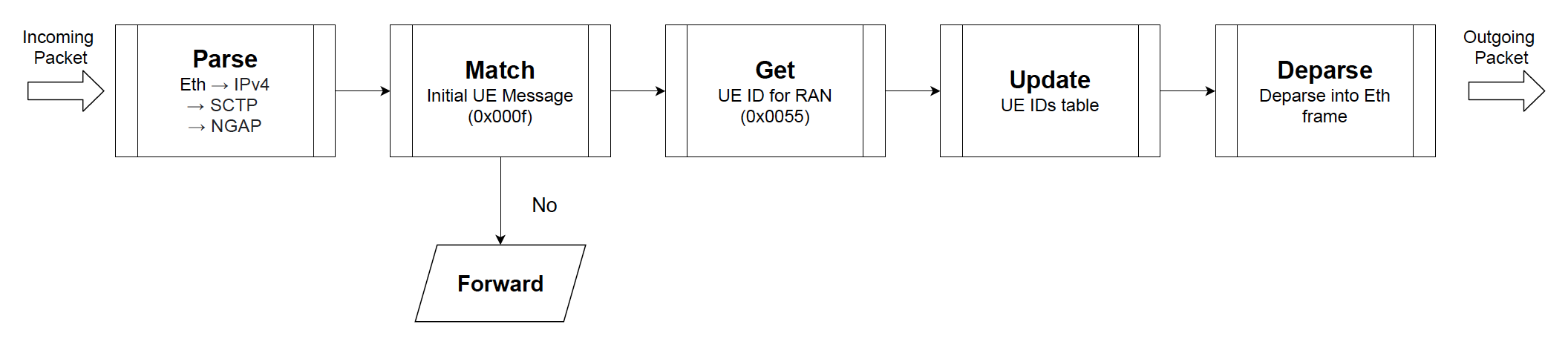}
    \caption{The NGAP pipeline between between gNBs and AMF as appeared in our work~\cite{Gokarslan2022}.}
    \label{fig:ngap-pipeline}
    
\end{figure*}

\section{The Cellular Data Plane}\label{sec:arch-dp}

This section details the packet processing pipeline at the cellular data plane aiming to achieve URLLC latency targets, specifically on the N3 link, as shown in Figure~\ref{fig:5g-arch}. In addition to latency gains, our design empowers cellular networks to achieve monitoring up to the application layer with enhanced network security. 

As next-generation cellular networks, including the 5G, target massive networks with increased in-network communication (e.g., industrial cellular networks), we aim to reduce the latency between users of the same cellular network. To this end, we propose a novel method, \textit{intra-cellular optimization}, that allows an authorized set of users to communicate without involving UPF with the packet transfer. Thanks to the programmable switches that interconnect the gNBs and cellular core, the switches forward the GTP packets from known UE pairs back to the respective gNB. Intra-cellular optimization, therefore, significantly reduces the latency of the UE pairs. It further reduces the load on UPF and consequently improves the overall network performance in terms of latency and throughput. UCP allows network engineers to manage the authorization set at the programmable switches in real-time while actively updating the table with the TEID from the GTP packets matched with UE IP addresses. Further, UCP runs a centralized shortest-path routing algorithm between switches to distribute paths with TEID data so that switches can forward GTP packets to switch connecting to the respective gNB.

In addition to intra-cellular optimization, we implement a fine-grained network monitoring mechanism on the programmable switch and utilize UCP to poll real-time monitoring statistics. Finally, we design a fine-grained firewall with either whitelisting or blacklisting behavior (depending on the decision of default forwarding behavior). We implement our design in P4 to support IPv4, ICMP, TCP, UDP, and SCTP, yet our implementation can further be extended to support different types of network protocols.

\begin{figure*}[!ht]
    \centering
    
    %columnwidth,keepaspectratio
    \includegraphics[width=\linewidth]{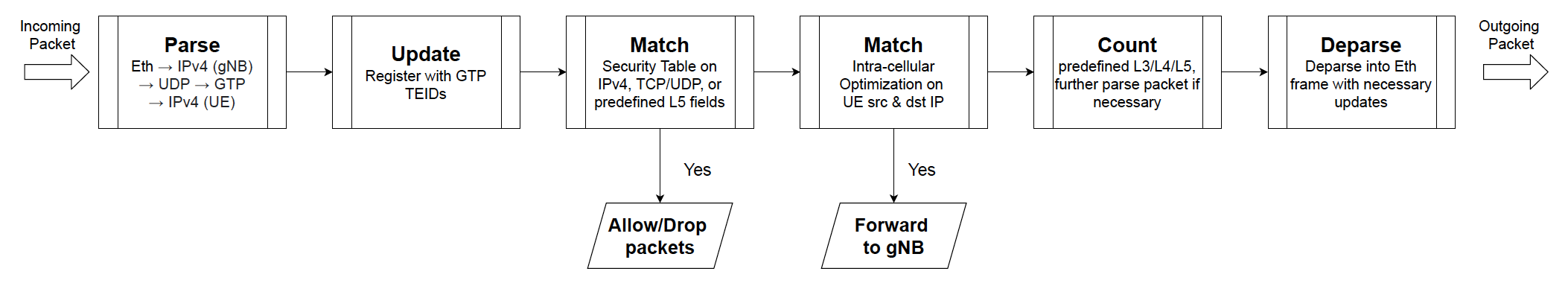}
    \caption{The pipeline parsing GTP packets on the N3 link as appeared in our preliminary work~\cite{Gokarslan2021}.}
    \label{fig:gtp-pipeline}
    
\end{figure*}

As shown in Figure~\ref{fig:gtp-pipeline}, our packet processing pipeline supports GTPv2 packets that encapsulate IPv4 packets from the UE. We keep a mapping between UE IP addresses with their respective GTP downstream TEID, where the downstream is the GTP packet from gNB to UE. We use this map to determine destination gNB on intra-cellular optimization. The pipeline then examines the packet in terms of the security rules and takes the necessary actions, such as dropping the packet. If the pipeline does not drop the packet due to security rules, it applies the intra-cellular network optimization table, keeping the authorized UE IP addresses. In case of a hit on this table, the pipeline modifies the packet into a GTP downstream, updates the TEID, and forwards it to gNB with the respective UE (or another switch that connects to that gNB). Before forwarding the packet, the pipeline applies the monitoring rules on the decapsulated IP packet. The pipeline finally encapsulates the packet back to GTP and forwards it accordingly.

\section{Mathematical Model of the Intra-Cellular Latency}\label{sec:arch-math}

In this section, we present the mathematical model that we design to evaluate the end-to-end latency of UE communication. We utilize our model in the evaluation section for the bigger cellular networks.

\begin{definition}[Node]
A node is either (a) UE, (b) gNB, (c) a switch, (d) UPF.
\end{definition}
\newpage
\begin{definition}[Network]
A network ($N$) is a graph of nodes, where the node types have the following limitations
\begin{enumerate}
    \item UE ($ue_i$), can only be the first or the last node in the graph
    \item gNB ($gnb_j$) can only be in between UE and a switch
    \item a switch ($sw_k$) cannot be connected to UE
    \item UPF ($u$) is one-and-only in a network, and it only connects to a switch
\end{enumerate}
\end{definition}

We use the following definitions from our preliminary work~\cite{Gokarslan2022}.

\begin{definition}[Link]
A link $l_{ij}$ is an undirected pair between two nodes ($n_i$ and $n_j$), where $l_{ij} = l_{ji}, \forall i, j \in \mathbb{N}$. A link has a latency value $l_{ij}.latency \in \mathbb{Q}^{\geq 0}, \forall i, j$.
\end{definition}
\begin{definition}[Path]
A path $P_{ij}$ of $ue_i$ and $ue_j$ is the smallest (in terms of number of unique nodes) sub-network $P_{ij} \subseteq N$ that connects $ue_i$ and $ue_j$ and contains the UPF node, that is, $u \in P_{ij}$ and $\forall u \in P_{ij}^\prime$ connecting $ue_i$ and $ue_j$, $\|P_{ij}\| \leq \|P^\prime_{ij}\|$.
\end{definition}

In addition to the path, we define an optimized path for intra-cellular optimization, as shown below.

\begin{definition}[Optimized Path]
An optimized path $P_{ij}$ of $ue_i$ and $ue_j$ is the smallest (in terms of number of unique nodes) sub-network $P_{ij} \subseteq N$ that connects $ue_i$ and $ue_j$ and does not contain the UPF node, that is, $u \notin P_{ij}$ and $\forall u \notin P_{ij}^\prime$ connecting $ue_i$ and $ue_j$, $\|P_{ij}\| \leq \|P^\prime_{ij}\|$.
\end{definition}

\begin{definition}[End-to-end Latency]
We define the path end-to-end latency ($L_{ij}$ of path $P_{ij} = \{ue_i, gnb_i^\prime, ..., gnb_j^\prime, ue_j\}$) as the sum of latency of each link in path $P_{ij}$. That is,
\begin{equation}
    L_{ij} = \sum_{(n_i^\prime, n_j^\prime) \in P_{ij}} l_{i^\prime j^\prime}.latency
\end{equation}
\vspace{1em}
where and $\forall l_{ij} = 0$ iff $n_i$ is UE or $n_j$ is UE.
\end{definition}
The end-to-end latency without intra-cellular optimization for an optimized path $P_{ij}$ is:
\begin{equation}
    l_p = gnb_i^\prime + gnb_j^\prime + S_{iu} + S_{uj} + u
\end{equation}
where 
$S_{ij}$ is the switch latency between nodes $n_i$ $n_j$.
Similarly, the latency equation becomes as follows for an optimized path:
\begin{equation}
    l_o = gnb_i^\prime + gnb_j^\prime + S_{ij}
\end{equation}
since the link to UPF from $ue_i$ and another connection back from the UPF to $ue_j$ are not in the the path.

\chapter{IMPLEMENTATION}\label{chapter:implementation}
This section gives the details of the implementation of our architecture that we introduced in Chapter~\ref{chapter:architecture}. The implementation contains two parts: (a) Emulation: we have implemented a single switch topology with two gNBs (each having a single UE) with an open-source 5G core, Open5GS. We run the P4 code on BMV2, a P4 software switch, and deploy each gNB and the 5G core in individual virtual machines. This implementation aims to evaluate and demonstrate our architecture's capabilities in a real 5G network and evaluate the latency gain. (b) Simulation: As we mentioned earlier, UCP can also run on multi-switch topologies. Due to the hardness of emulating multiple P4 software switches, we implement a Python simulator that simulates the switch behavior. We use this simulation implementation to evaluate the UCP under different switches and gNBs.

\section{Emulation}\label{sec:implementation-emulation}
\vspace{1em}
\begin{figure}[H]
    
    \centering
    \includegraphics[width=0.95\linewidth]{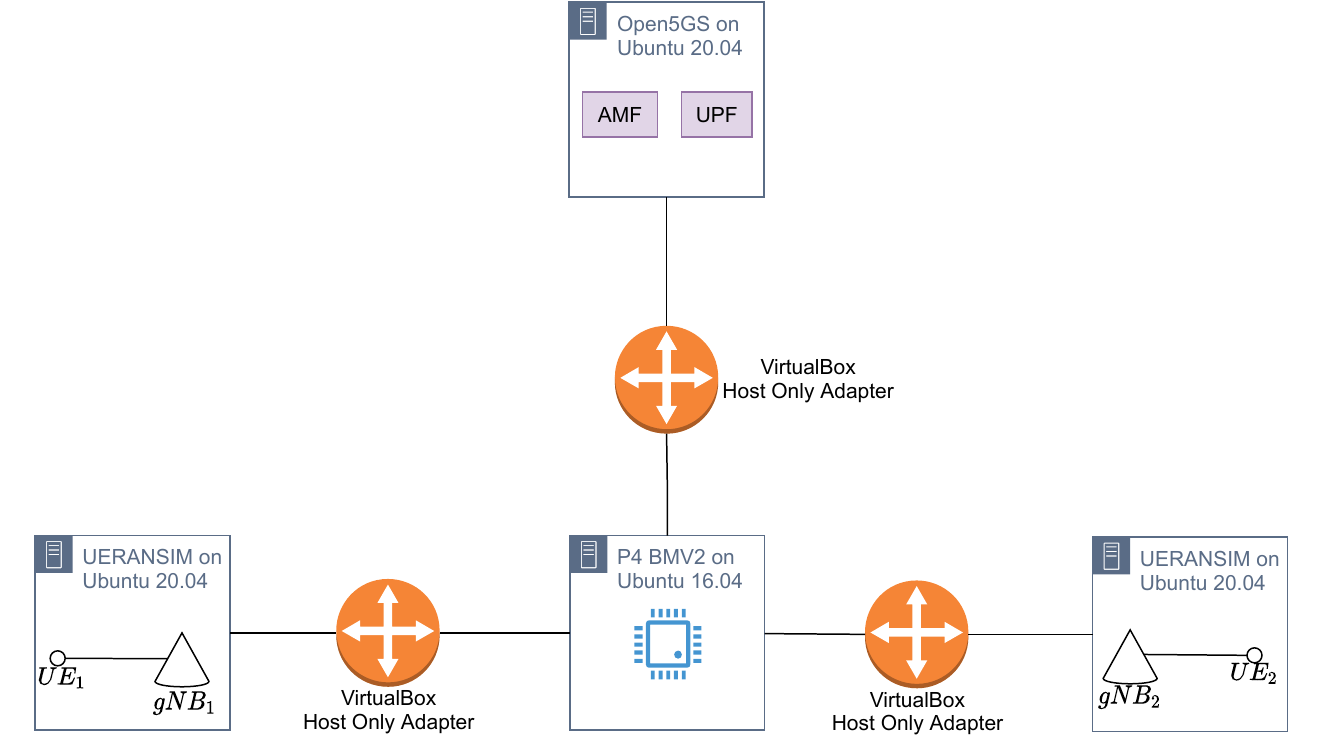}
    \caption{Emulation topology with VirtualBox~\cite{virtualbox}. Table~\ref{table:implementation-vm-spec} shows the specification of each VM in the topology.}
    \label{fig:implementation}
    
\end{figure}

In the first part of our implementation, we design a 2-UE topology to demonstrate our programmable data path on a full 5G stack. We pick the P4's behavioral model (BMV2)~\cite{bmv2} software switch to run our pipeline written in P4 version 16 (P4$_{16}$~\cite{p416}). We pick Open5GS~\cite{open5gs} for 5G stack that provides core network functions and UERANSIM~\cite{ueransim} as the UE and RAN simulator, which provides RAN implementation including UE and gNBs. As depicted in Figure~\ref{fig:implementation}, we deploy two virtual machines (VM) where each runs a UE and gNB, another VM running Open5Gs, and one final VM that interconnects other VMs and deploys the P4 software switch on a MacBookPro with 16 GB RAM and an Intel i7 (4-cores) clocked at 2.9 GHz. We use Host-Only Adapter~\cite{virtualboxhostonly} as the network adapter between each VM. We use Ubuntu Server 20.04~\cite{ubuntu} as the OS in each VM, except for the VM running P4, which has Ubuntu Server 16.04.

As our implementation works on top of software emulation, we evaluate the gain of intra-cellular optimization compared to traditional cellular network usage on the same topology. This is because BMV2 is instead a development tool that is not designed as a production-grade software switch, per its documentation. It is, therefore, logical to expect higher latency gains on a hardware P4 switch such as Tofino.

\begin{table}[htbp]
\vspace{1em}
    \caption[The specification of each VM in our evaluation.]{The specification of each VM in our evaluation. We have distributed all available resources (4 vCPU Cores) and near 10 GB of RAM with respect to the workload of each VM after running a couple of experiments with different sets of specifications.}
    \vspace{0.5em}
\begin{center}
    \begin{tabular}{|c|c|c|c|}
            \hline
            VM & OS & v-CPU Cores & RAM \\ \hline \hline
            Open 5Gs & Ubuntu 20.04 & 1 & 3072 MB \\ \hline
            UERANSIM ($gNB_1$/$gNB_2$) & Ubuntu 20.04 & 1 & 2048 MB \\ \hline
            P4 BMV2 & Ubuntu 16.04 & 2 & 4096 MB \\ \hline
    \end{tabular}
    \label{table:implementation-vm-spec}
    
\end{center}
\end{table}

\subsection{P4 Implementation} 
\begin{figure}[htbp]
    \centering
    \includegraphics[width=0.9\linewidth]{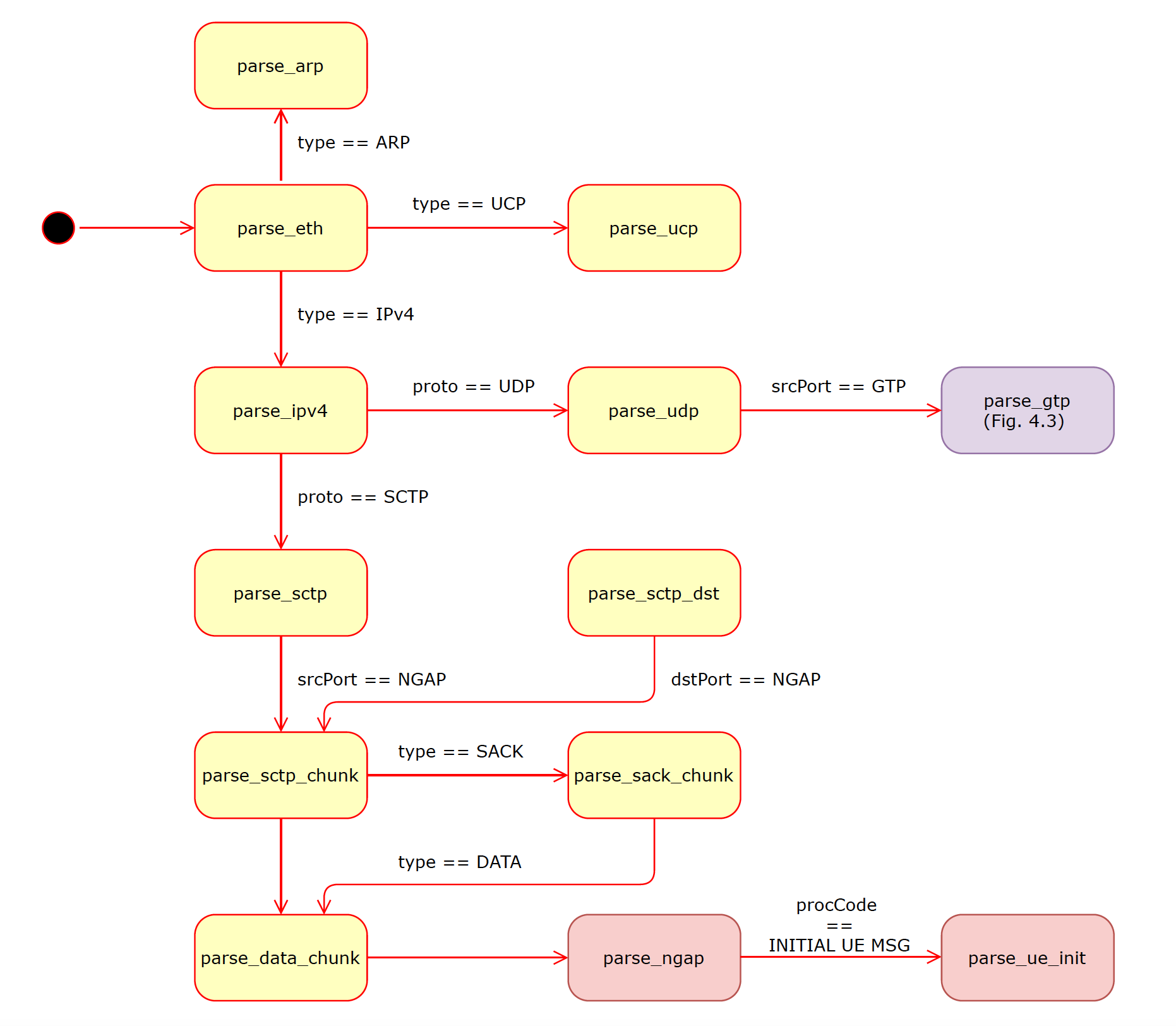}
    \caption{FSM diagram for header parsing. At each state, a protocol layer is decapsulated. States after parse\_gtp is shown in Figure~\ref{fig:p4-fsm-gtp}. Initially appeared in~\cite{Gokarslan2022}.}
    \label{fig:p4-fsm}
    
\end{figure}

\begin{figure}[htbp]
    \centering
    \includegraphics[width=0.9\linewidth]{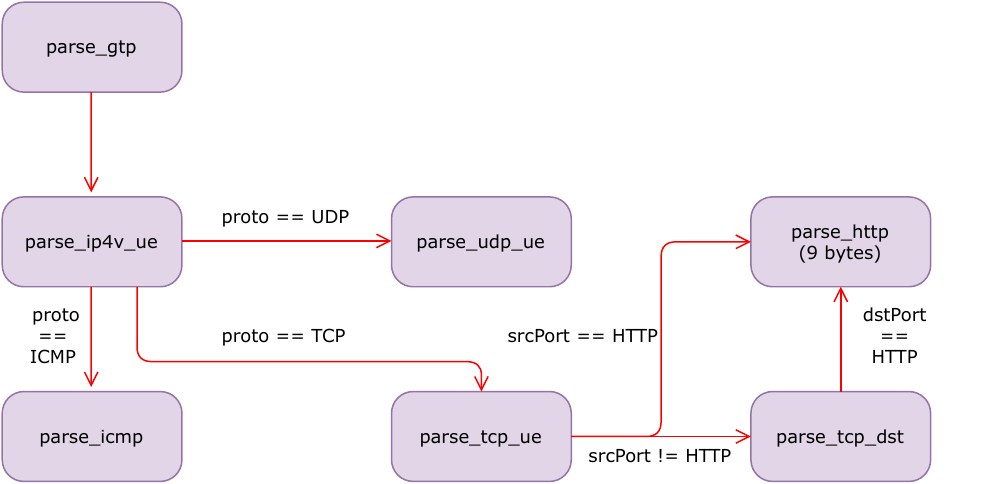}
    \caption[P4 header parsing finite state machine diagram for GTP packets.]{P4 header parsing finite state machine diagram for GTP packets. HTTP port is 80~\cite{rfc7230}. Only 9-byte HTTP packets are parsed at the parse\_http state as P4 only allows constant header size.}
    \label{fig:p4-fsm-gtp}
    
\end{figure}

Our programmable data plane contains two parts: (a) A fine state machine (FSM) that parses the incoming packet's headers as shown in Figure~\ref{fig:p4-fsm} and Figure~\ref{fig:p4-fsm-gtp}, and (b) packet processing algorithm. We first parse the L2 (i.e., Ethernet) header and continue until the application layer (i.e., GTP and NGAP), depending on the content. For the cellular data packets (i.e., GTP), we also parse up to the IPv4 packet encapsulated into the GTP. Note that parsing of the encapsulated IPv4 packet can be modified depending on the user's needs. The parsing FSM we use in our evaluation is shown in Figure~\ref{fig:p4-fsm-gtp}.

P4 program runs the processing algorithm utilizing the headers that are parsed in the first stage as shown in Figure~\ref{alg:p4-algorithm} after the packet header processing. 
The algorithm applies tables we define in Table~\ref{table:p4-tables} in order to update P4 registers (e.g., for counting) depending on the packet content. Note that our P4 program has an L2-switch behavior (i.e., MAC-based frame forwarding) for other types of packets as shown in Figure~\ref{fig:packet-flowchart}.

\makeatletter
\setlength{\@fptop}{0pt}
\makeatother

\makeatletter
\setlength{\@fptop}{0pt}
\makeatother

\begin{figure}[H]
\label{alg:p4-algorithm}
\begin{center}
\framebox[5.8in]{
\begin{minipage}[t]{5.9in}
\begin{algorithmic}
\IF {ipv4\_ue}
    \STATE apply table ipv4\_downstream\_teid
    \STATE apply table ipv4\_in\_network
    \STATE apply table ipv4\_\{in, out\}\_\{wlist, blist\}
    \STATE apply \{tcp, udp\}\_\{in, out\}\_ \\ \{wlist, blist\}
    \IF {tcp\_ue}
        \STATE apply http\_sensor and increment related counter
    \ENDIF
\ELSIF{ngap\_init}
    \STATE add ngap\_init.value to ue\_ids list
\ELSIF {ucp}
    \STATE match on ucp.opId and take the related action in Table~\ref{tab:p4control-operations}
\ENDIF
\end{algorithmic}
\end{minipage}
}
\end{center}

\caption[Pseudocode of the main processing algorithm implemented in P4.]{Main processing algorithm implemented in P4. Note that for all other types of frames, we have implemented a L2-switch behavior.}
% \vskip
\end{figure}
\section{Multi-Switch Simulation}\label{sec:implementation-multisw}
As our emulation implementation requires high computing resources, we develop a Python simulation to evaluate the performance of our system on larger topologies having multiple P4 switches. Our simulation implementation consists of four parts, namely: (a) topology generation, (b) route computation and TEID announcement, (c) UCP implementation, and (d) intra-cellular latency simulation.

\begin{table}[thbp]
\vspace{0.5em}
\caption[Tables we define in the P4 program.]{Tables we define in the P4 program. We create a separate table for each name in parentheses, such as ipv4\_in\_blist.}
\vspace{0.5em}
\begin{center}
    \begin{tabular}{|c|c|c|}
            \hline
            Table & Match\\ \hline \hline
            ngap\_ue & ngap\_ue\_id  \\ \hline
            ipv4\_down\_teid & \makecell{ipv4.src, ipv4\_ue.dst (lpm\tablefootnote{longest prefix match: matching with the most specific address~\cite{lpm}.})}  \\ \hline
            ipv4\_in\_network & \makecell{ipv4.src, ipv4\_ue.dst (lpm)}  \\ \hline
            ipv4\_\{in, out\}\_\{wlist, blist\} & ipv4.\{src, dst\} (lpm)  \\ \hline
            \makecell{\{tcp, udp\}\_\{in, out\}\_ \\ \{wlist, blist\}} & \{tcp, udp\}.\{srcPort, dstPort\}  \\ \hline
            http\_sensor & http\_ue.command \\ \hline
    \end{tabular}
    \label{table:p4-tables}
    
\end{center}
\end{table}

% \begin{table}[H]
% %\begin{center}
%     \caption[Tables we define in the P4 program.]{Tables we define in the P4 program. We create a separate table for each name in parentheses, such as ipv4\_in\_blist.}
%     \begin{tabular}{|c|c|c|}
%             \hline
%             Table & Match & Description \\ \hline \hline
%             ngap\_ue & ngap\_ue\_id & \makecell{Initiating UE} \\ \hline
%             ipv4\_down\_teid & \makecell{ipv4.src, ipv4\_ue.dst (lpm\tablefootnote{longest prefix match: matching with the most specific address~\cite{lpm}.})} & \makecell{GTP packets coming\\ from UPF to gNBs.}   \\ \hline
%             ipv4\_in\_network & \makecell{ipv4.src, ipv4\_ue.dst (lpm)} & \makecell{Intra-cellular optimization\\ for known UE.}   \\ \hline
%             ipv4\_\{in, out\}\_\{wlist, blist\} & ipv4.\{src, dst\} (lpm) & \makecell{For whitelisting or blacklisting\\ at IP address level.}   \\ \hline
%             \makecell{\{tcp, udp\}\_\{in, out\}\_ \\ \{wlist, blist\}} & \{tcp, udp\}.\{srcPort, dstPort\} & \makecell{For whitelisting or blacklisting\\ at TCP and UDP levels.}   \\ \hline
%             http\_sensor & http\_ue.command & \makecell{For incrementing \\counter for the command.} \\ \hline
%     \end{tabular}
%     \label{table:p4-tables}
    
% %\end{center}
% \end{table}

\subsection{Topology Generation}

We randomly generate topologies with gNBs between up to 200 and 40 switches at maximum. Each gNB has at most 10 devices. 

The UPF and AMF connect to a topology via separate P4 switches, as shown in Figure~\ref{alg:topology}. An example topology generated by this algorithm is shown in Figure~\ref{fig:multisw-example}. We vary the switch-to-gNB ratio to evaluate the effect of the switch-to-gNB ratio on the end-to-end latency.

\begin{figure}[H]
\label{alg:topology}
\begin{center}
\framebox[5.8in]{\begin{minipage}[t]{5.9in}
\begin{algorithmic}
\STATE Given switch-to-gNB ratio ($G$), number of switches ($S$), and maximum \\ gNB-to-UE ratio ($U$).
\STATE Use $\textbf{nodes}$ an array of Node
\begin{ALC@g}
    \STATE nodes $:= \{UPF, AMF, SW_i, gNB_{ij}, UE_{ijk_{ij}}\}$ \\
        \hspace{4em} where $i \leq S, j \leq G, k_{ij} \leq K_{ij}=$uniform($1, U$)
    \STATE connect ($SW_1$, $UPF$) and ($SW_2$, $AMF$) pairs
    \STATE connect $SW_i$ and $gNB_{ij}$, $\forall i \leq S, j \leq G$
    \STATE connect $gNB_{ij}$ and $UE_{ijk_{ij}}$, $\forall i \leq S, j \leq G, k_{ij} \leq K_{ij}$
    \STATE $E$ = uniform($s / 2, s - 1$)
    \STATE connect $SW_i$ and $SW_j$ for a total of $E$ randomly \\ generated unique pairs of $(i, j)$
\end{ALC@g}

\end{algorithmic}
\end{minipage}}
\end{center}
\caption[Topology Generation Algorithm.]{Topology generation algorithm simulation.}
% \vskip
\end{figure}

\begin{figure}[H]
    
    \centering
    \includegraphics[width=\linewidth]{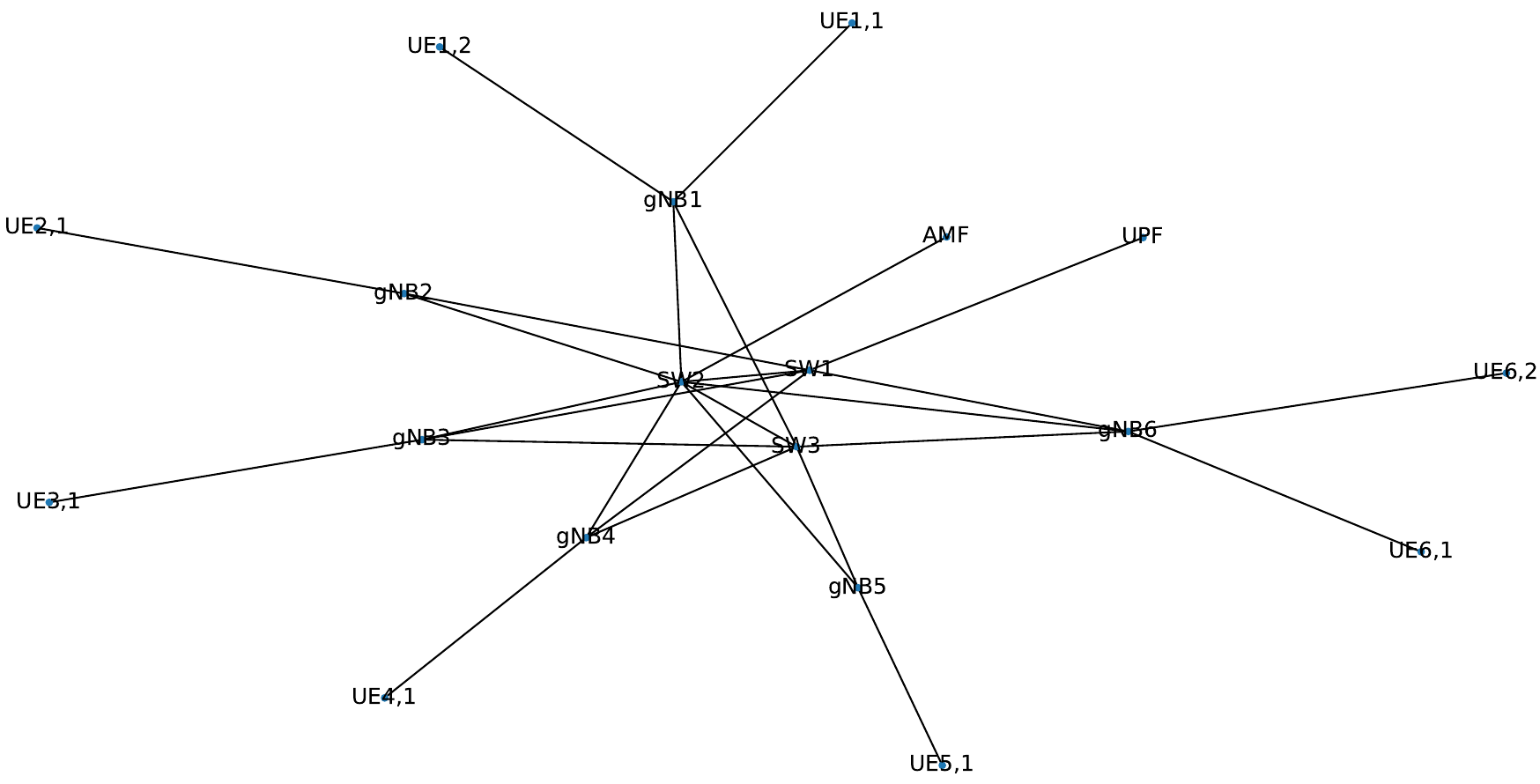}
    \caption[A randomly generated topology with 6 gNBs and 3 P4 switches.]{A randomly generated topology with 6 gNBs and 3 P4 switches where a gNB has at most 2 UE. UPF is connected to topology via SW1 while AMF is connected via SW2. We generate this topology with the algorithm in Figure~\ref{alg:topology}.}
    \label{fig:multisw-example}
    
\end{figure}

\subsection{Route Computation and TEID Announcement}
In the first part of our simulation, we implement the simulated behavior of each switch. We implement a Link State Routing (LSR) algorithm, inspired by the OSPF~\cite{rfc2328}, between switches where each switch announces its link-state periodically. Similar to the routing, we have also implemented a TEID announcement algorithm where each switch announces the updates on its TEID table instantaneously. We evaluate the TEID distribution behavior in Section~\ref{ssec:eval-teid}.

% \begin{figure}[H]
% \label{alg:teid-distribution}
% \begin{center}
% \framebox[6.0in]{\begin{minipage}[t]{5.9in}
% \begin{algorithmic}
% \STATE Given number of gNBs per switch ($G$), switches ($S$), and max UE per gNB ($U$).
% \STATE Use $\textbf{nodes}$ an array of Node
% \STATE \kerim{check the algorithm!!!}
% \begin{ALC@g}
%     \STATE nodes $:= \{UPF, AMF, SW_i, gNB_{ij}, UE_{ijk_{ij}}\}$ \\
%         \hspace{4em} where $i \leq S, j \leq G, k_{ij} \leq K_{ij}=$uniform($1, U$)
%     \STATE connect ($SW_1$, $UPF$) and ($SW_2$, $AMF$) pairs
%     \STATE connect $SW_i$ and $gNB_{ij}$, $\forall i \leq S, j \leq G$
%     \STATE connect $gNB_{ij}$ and $UE_{ijk_{ij}}$, $\forall i \leq S, j \leq G, k_{ij} \leq K_{ij}$
%     \STATE $E$ = uniform($s / 2, s - 1$)
%     \STATE connect $SW_i$ and $SW_j$ for a total of $E$ randomly generated unique pairs of $(i, j)$
% \end{ALC@g}

% \end{algorithmic}
% \end{minipage}}
% \end{center}
% \caption[TEID Distribution Algorithm.]{TEID Distribution Algorithm on each switch.}
% % \vskip
% \end{figure}

\subsection{UCP Implementation}
We run a process for each switch in the topology using the Python's \\\texttt{multiprocessing}~\cite{multiprocessing} library. We further execute another process that runs as the UCP controller. Each switch process runs a simulation version of the P4 main loop we describe above, where a switch process can (a) receive and create UCP messages, (b) process data messages from and to the other switches. 

\subsection{Intra-Cellular Latency Evaluation Simulation}

Due to computational resource limitations, we design our emulation platform with two pairs of gNB-UE. On the other hand, real-world cellular applications can contain many more devices and gNBs, requiring multiple programmable switches to interconnect them. To demonstrate our design's performance on a large-scale topology, we implement a Python simulation of the switches where each switch runs the algorithms in Figure~\ref{alg:multiswitch} as a process based on Python's \texttt{multiprocessing} library. We then implement another process that is responsible for the UCP and based on the algorithms in Figure~\ref{alg:multiswitchucp}. We use the mathematical model of latency in Section~\ref{sec:arch-math} to normalize the latency values we gather from the simulation.

\chapter{EVALUATION}\label{chapter:evaluation}
In this section, we first present the evaluation results using the emulation defined in Section~\ref{sec:implementation-emulation}. With our emulation platform, we evaluate (a) latency improvement with the intra-cellular optimization, (b) NGAP processing at the data plane using P4, (c) in-network security capabilities, and (d) fine-grained network monitoring. We, then, discuss the evaluation results with our simulation focusing on multi-switch topologies: (a) TEID announcement and route computation with UCP, and (b) latency improvement with intra-cellular optimization. We finally conclude this chapter with a comparison between latency results and the mathematical model of latency described in Section~\ref{sec:arch-math}.

\section{Evaluation with Cellular Network Emulation}
\begin{figure}[!ht]
    \centering
    
    \includegraphics[width=\columnwidth,keepaspectratio]{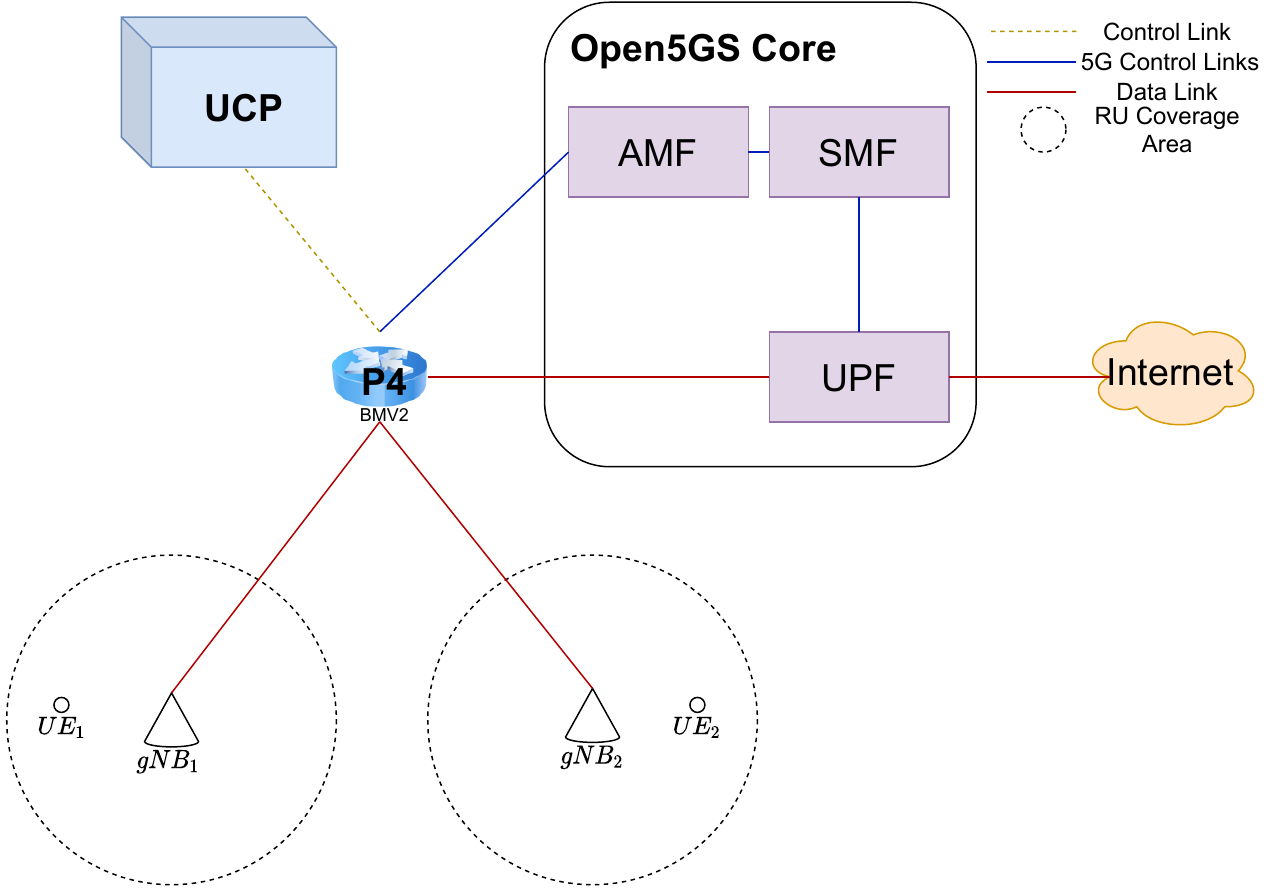}
    \caption[The emulation architecture with two gNBs.]{The emulation architecture with two gNBs ($gNB_i$) with the same signal coverage (shown in circles). UCP manages the P4 switch in real-time, and the switch processes both NGAP and GTP packets.}
    \label{fig:emulation-arch}
    
\end{figure}

Figure~\ref{fig:emulation-arch} shows that we use a topology with a single P4 switch that connects the gNBs to each other and the 5G cellular network. We describe the details of the implementation in Section~\ref{sec:implementation-emulation}. We use the same topology for all experiments in this section except for the last one, where we define a slightly different topology that includes connection to the outside of the cellular network.

\subsection{Intra-Cellular Latency Evaluation} 

The devices on a cellular network use a dedicated network function (e.g., UPF in 5G) to reach to the outside networks (e.g., the Internet). Moreover, this requirement is still valid for two users in the same network, as the RAN passes data path packets (tunneled using GTP) between UE and respective cellular function. Unfortunately, this creates an additional delay for the users in the same cellular networks. With the emergence of the newer use cases for cellular networks such as industrial networks, this problem further exacerbates since many devices in such networks communicate with other devices within the network. 

\begin{figure}[!ht]
    \centering
    \includegraphics[width=\linewidth]{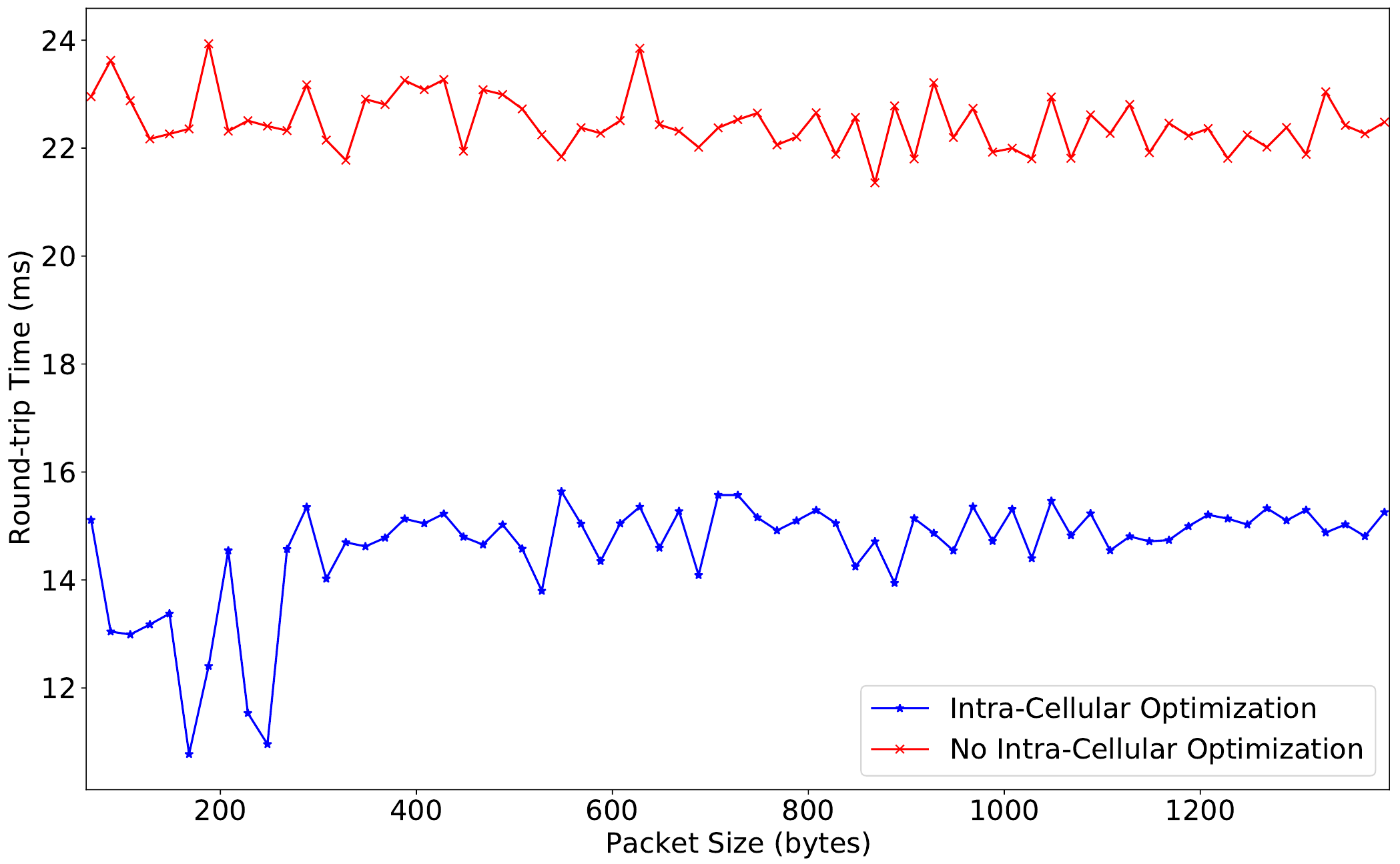}
    \caption{Average RTT evaluation between $UE_1$ and $UE_2$ with \texttt{ping}~\cite{ping} tool. We take an average of 100 runs for each packet size. Intra-cellular optimization always performs better than the traditional 5G setup and can achieve up to 2.2x fewer latency values. The results indicate no correlation between packet sizes and latency.}
    \label{fig:latency-result}
    
\end{figure}

Moreover, network administrators in such private networks have control over the devices (e.g., IIoT sensors in an industrial cellular network); therefore, they can control or pre-authorize the devices for network usage and make session management controls for such devices "unnecessary". With such considerations in mind, we design architecture in programmable switches where we utilize a TCAM table to hold the IP addresses and GTP TEIDs to forward GTP packets from known UE to the relative gNB instead of sending them to the UPF. We discuss the details of our design in Section~\ref{sec:arch-dp}. As shown in Figure~\ref{fig:latency-result}, our design can decrease latency by a factor of two in most of the cases with an average of 1.5x. In addition, our experiment shows that our system can work on packet sizes smaller than the Ethernet's conventional MTU of 1500 bytes. This limitation can be avoided by modifying our implementation to handle IP packets spanned to multiple Ethernet Frames. An earlier version of this experiment appeared in our preliminary work~\cite{Gokarslan2021}.

\makeatletter
\setlength{\@fptop}{0pt}
\makeatother

\subsection{NGAP Processing}
In addition to the GTP processing pipeline, we design an NGAP processing pipeline at the programmable data plane to parse the NGAP packets. It is relatively simple, matching initial UE messages and storing the UE IDs in a P4 register. Our design is a proof-of-concept demonstrating a complex protocol such as NGAP could be processed at the switch at line speed. Consequently, in this experiment, we demonstrate the scenario where numerous UE registers send the initial UE message, and then we poll the registered UE IDs using UCP. 

\newpage 

Table~\ref{table:ngap-result} shows the average register retrieval time via UCP with respect to poll interval, where we run 500 experiments for each poll interval. We see that the average interval time is between 29 - 68 ms, and the higher poll interval causes the higher average time. 

\begin{table}
\caption{Average progressing time of programmable switch on UE ID requests on a topology containing four devices and two gNBs. We execute a total of 500 experiments on each poll frequency and take the average processing time.}
\vspace{0.5em}
\begin{center}
    \begin{tabular}{|c|c|}
            \hline
            Poll Frequency & Average Processing Time (ms) \\ \hline \hline
            0.1 Hz & 67.4 \\ \hline
            1 Hz & 65.6 \\ \hline
            10 Hz & 63.4 \\ \hline
            100 Hz & 57.2 \\ \hline
            1000 Hz & 47.2  \\ \hline
            $\infty$ Hz & 29.2 \\ \hline 
    \end{tabular}
    \label{table:ngap-result}
\end{center}
\end{table}

\subsection{Security Evaluation} \label{ssec:eval-security}
One aspect of our framework is network security. To this end, we implement a network firewall dedicated to UE traffic tunneled via GTP. The firewall uses the TCAM tables of P4 and is updated via UCP in real-time. While our design works at the line speed, it reduces the hardware cost of cellular networks as it can the traditional firewalls.

In this experiment, we evaluate the performance of our UCP implementation and P4-software switch. We update the security rules and measure the rule update time considering the total number of rules on the switch. We measure the rule update time from the UCP input until the rule's effect is seen on the dataplane. 

\newpage 

\begin{figure}[!ht]
    \centering
    \includegraphics[width=0.9\linewidth]{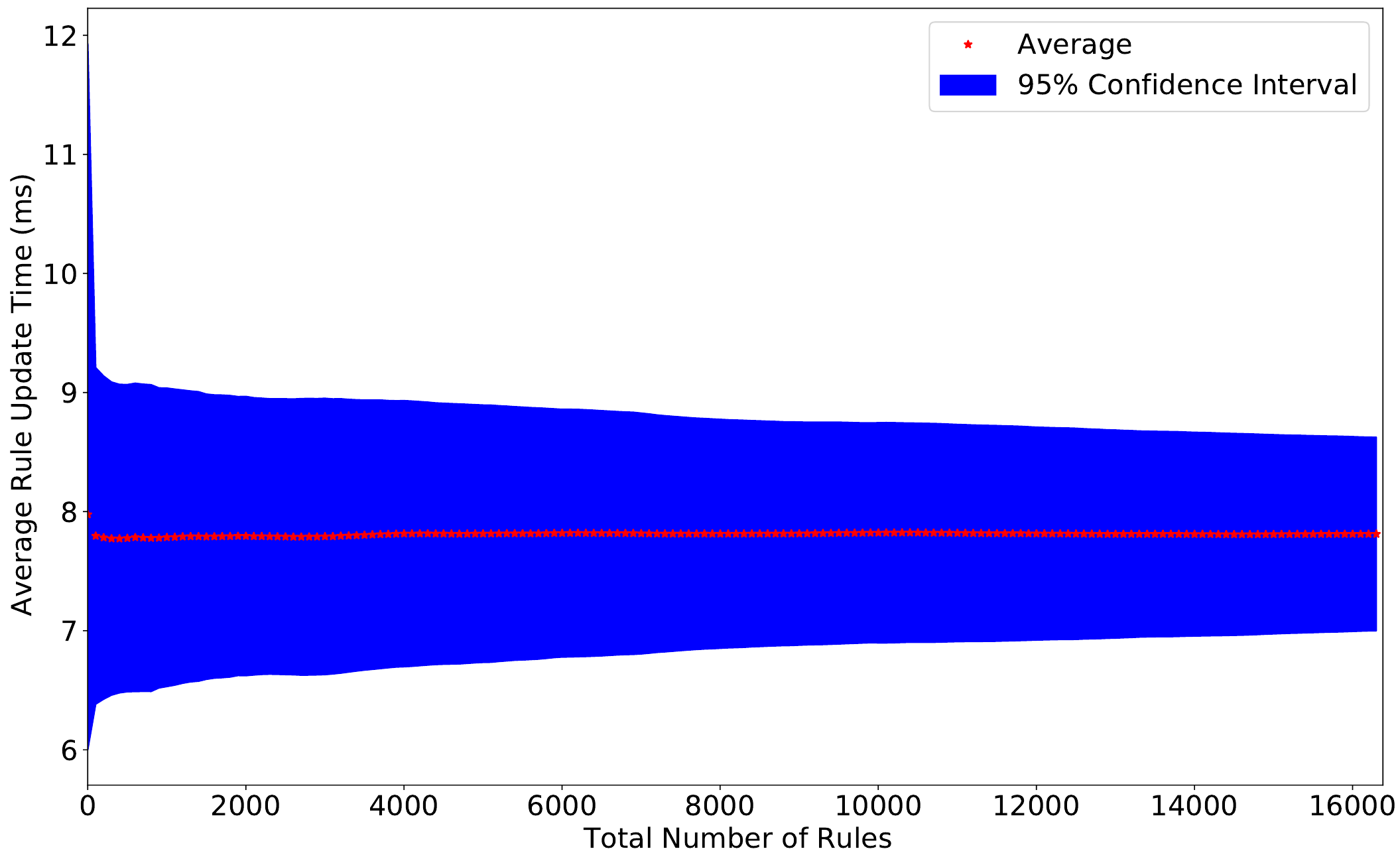}
    \caption{Average security rule update time with the total number of security rules on the P4 switch. Our results indicate no significant effect of the total number of rules on the rule update time, and, we can achieve less than 10 ms for each rule to be deployed on the dataplane with the P4 BMV2 software switch.
    }
    \label{fig:security-result}
    
\end{figure}

We implement a Python script to create HTTP servers with incremental port numbers, and it marks as the rule update when the connection to the server is broken, as we send "deny" rules with the specific port numbers. As shown in Figure~\ref{fig:security-result}, We found no significant evidence indicating a relation between the total number of rules and average rule update time. We measure the average rule update time less than 10 ms when the total number of rules exceeds 100 with a 95\% confidence interval.

\subsection{Monitoring Evaluation}\label{ssec:eval-monitoring}
\begin{figure}[!t]
    \centering
    \includegraphics[width=\linewidth]{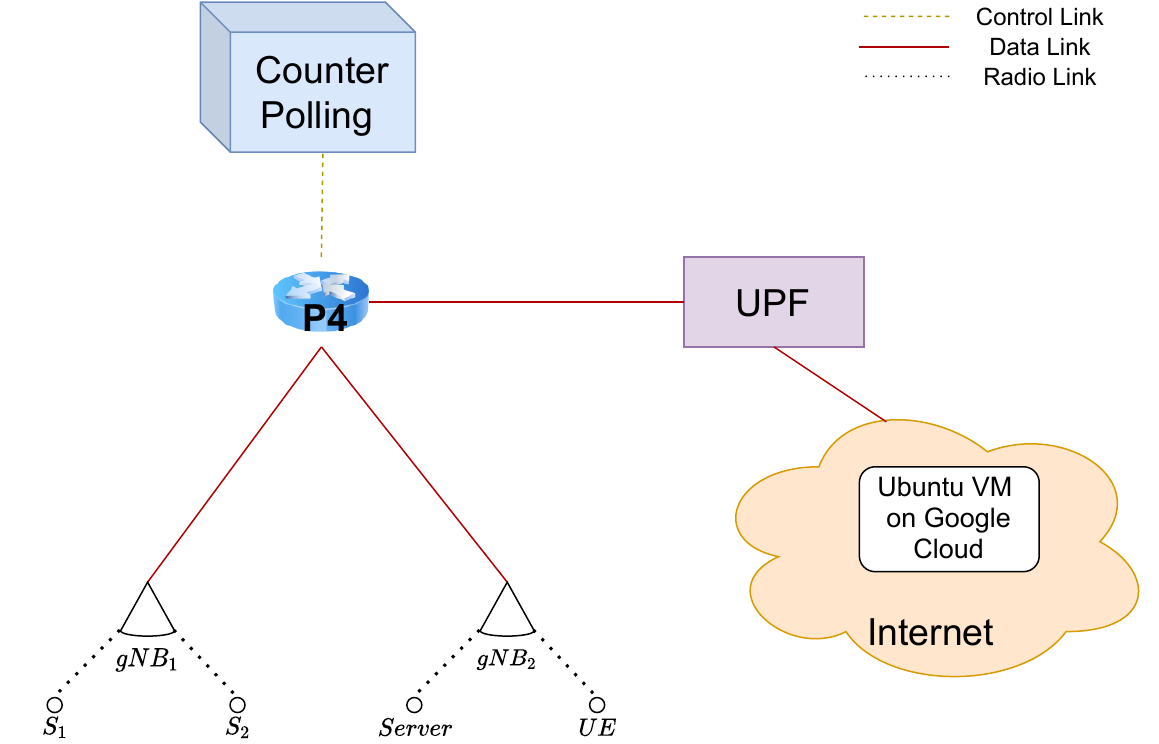}
    \caption{Two IoT sensors ($S_1$ and $S_2$) send heartbeat signals with a 10 ms period via HTTP to $Server$. The packet counter mechanism increments the counter for the respective sensor. In addition to the sensors, we run a live TCP traffic between the VM on Google Cloud and the $UE$, shown in the figure using \texttt{iPerf3}~\cite{iperf3}.}
    \label{fig:monitoring-architecture}
\end{figure}

This experiment adds IP, TCP, and HTTP counters for UE traffic via GTP. With the HTTP counter, we evaluate a scenario with IoT sensors. Each sensor sends a 2-byte heartbeat signal containing their ID. Our HTTP counter counts the number of heartbeat signals for each sensor, and we poll the counter status of each sensor via UCP. We also consider a TCP traffic with \texttt{iPerf3}~\cite{iperf3} between a server in Google Cloud and the UE, where the UE and IoT sensors are on the same cellular network. We show the whole evaluation topology in Figure~\ref{fig:monitoring-architecture}. 

We measure the throughput of the traffic between the UE and the server with respect to the polling interval of the HTTP counter via UCP. Table~\ref{table:monitoring-result}, shows that the traffic is significantly affected for less than 10 ms polling intervals. 

As we perform our evaluation on the BMV2 software switch, which runs on the same machine running the 5G network, the results with polling intervals shorter than 10 ms are insignificant. Further, a 10 ms polling interval for such a counter is unrealistic as many IoT devices have heartbeat intervals much higher than 10 ms.

\begin{table}[htbp]
\caption{TCP throughput between the $UE$ and VM with respect to packet counter polling intervals on the P4 switch via the UCP, as appeared in our preliminary work~\cite{Gokarslan2021}.  We run each \texttt{iPerf3} session five times on a pooling frequency of up to 10000 Hz.}
\vspace{0.5em}
\label{table:monitoring-result}
\begin{center}
    \begin{tabular}{|c|c|}
            \hline
            UCP Poll Frequency & Throughput \\ \hline \hline
            no polling (0 Hz) & 100.0\% \\ \hline
            1 Hz & 98.8\% \\ \hline
            10 Hz & 96.4\% \\ \hline
            100 Hz & 93.3\% \\ \hline
            1000 Hz & 89.1\% \\ \hline
            10000 Hz & 84.2\% \\ \hline
            no interval ($\infty$ Hz) & 74.5\% \\ \hline 
    \end{tabular}
\end{center}
\end{table}

\newpage

\section{Multi-Switch Simulation}
\begin{figure}[H]
    \centering
    \includegraphics[width=\linewidth]{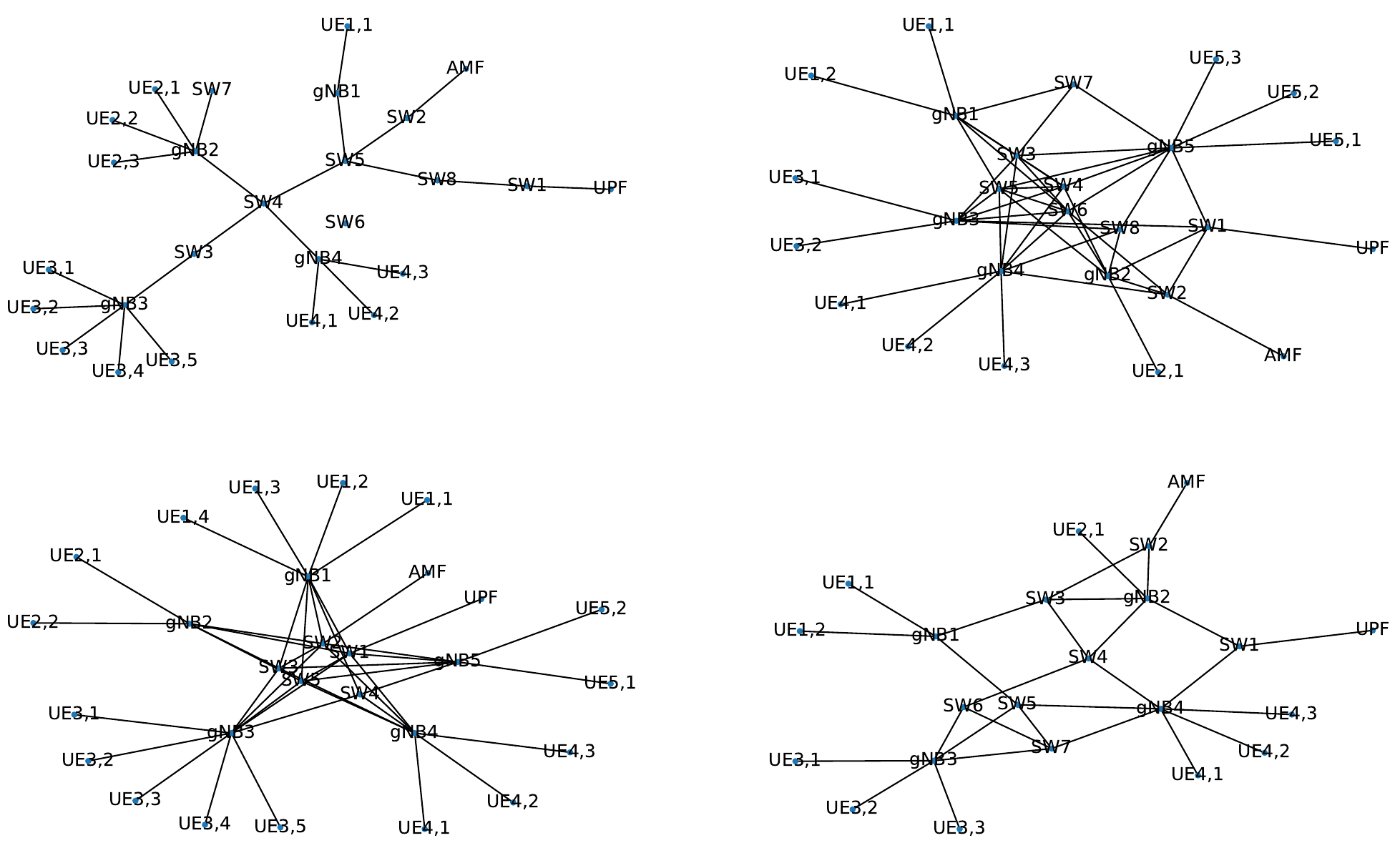}
    \caption{Some of the randomly generated topologies used in the evaluation.}
    \label{fig:multisw-topos}
    
\end{figure}

Due to resource constraints, we evaluate the multi-switch topology with a higher number of gNBs by implementing a Python simulation. The simulation runs for any arbitrary topology where $n$ P4 switches connect $m$ gNBs to AMF and UPF. We evaluate (a) the route computation and TEID advertisement performance at switches, and (b) the end-to-end communication latency with intra-cellular optimization and without it. Figure~\ref{fig:multisw-topos} shows some of the network topologies we use in the experimentation. Note that all of the topologies are randomly generated using Algorithm~\ref{alg:topology}.

\subsection{Route Computation and TEID Announcement}\label{ssec:eval-teid}
In this experiment, we test the implementation of TEID announcement algorithms that are deployed at the programmable switches shown in Figure~\ref{alg:multiswitch}.

We see a direct proportion between the number of switches and average retrieval and announcement durations; the larger network means a higher average distance between any pair of nodes. The results, therefore, suggest 250 ms or smaller average advertisement and retrieval times on networks containing 20 or fewer switches. For example, in a border case network with 100 switches, the average advertisement time reaches 1500 ms while the average retrieval time is around 500 ms.

\begin{figure}[H]
    \centering
    \includegraphics[width=\linewidth]{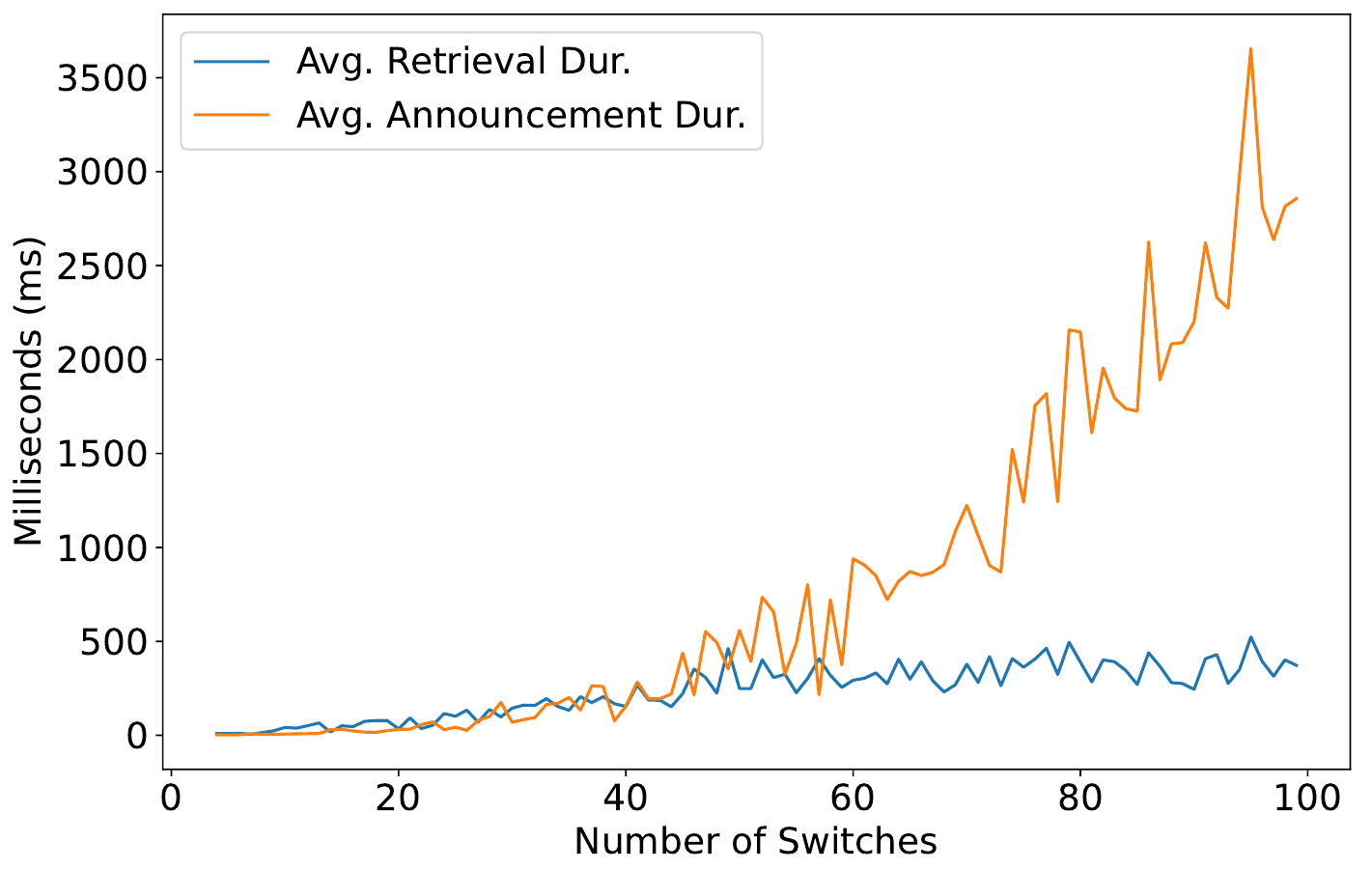}
    \caption{TEID advertisement and retrieval results are generated with the Python simulation. The topologies have fixed a switch-to-gNB and gNB-to-UE ratio of 5.}
    \label{fig:teid-durations}
    
\end{figure}

\subsection{Intra-Cellular Latency Evaluation on Simulation}
\begin{figure}[H]
    \centering
    \includegraphics[width=\linewidth]{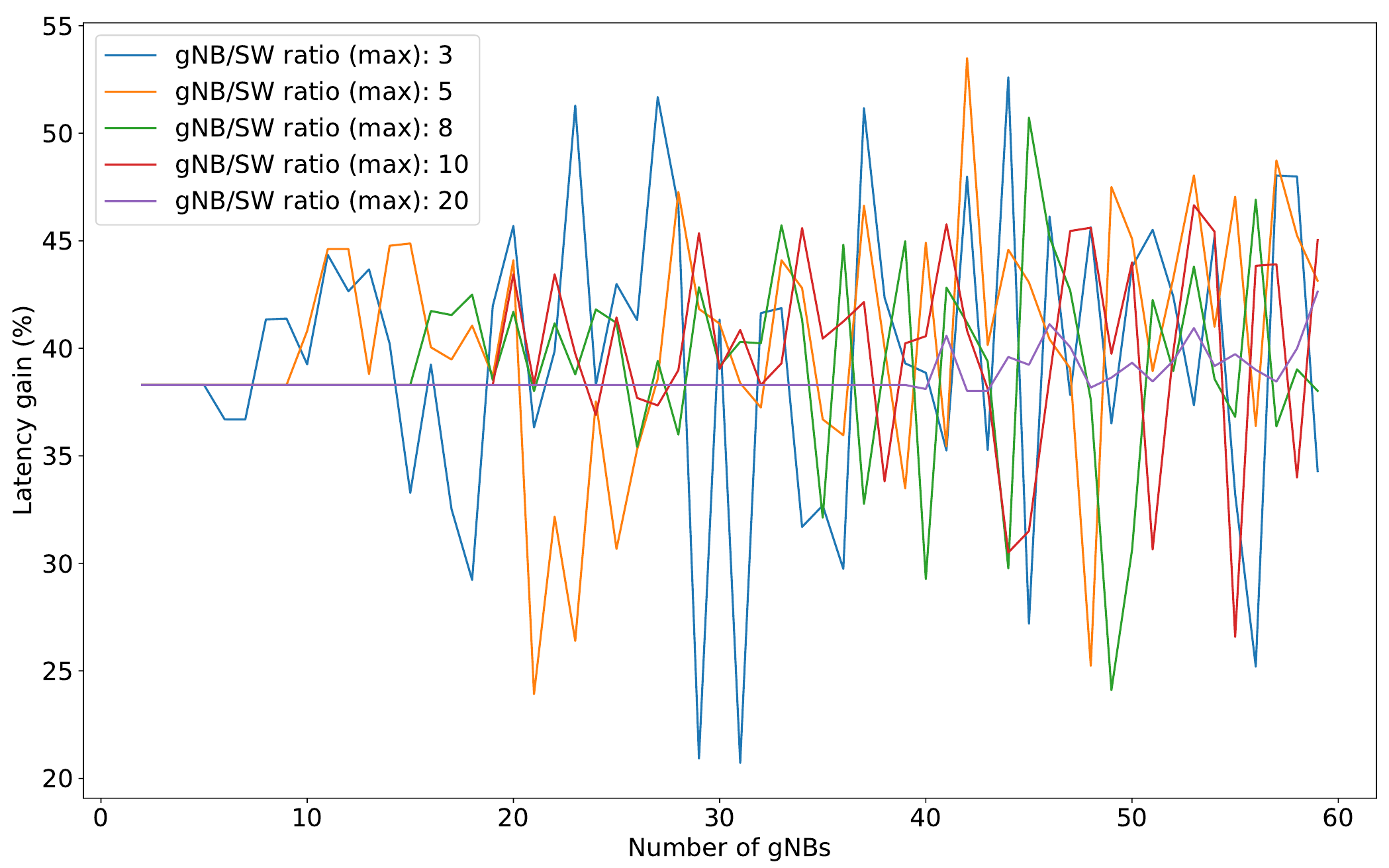}
    \caption{The lines in the figure show the average latency gain percentage for the different numbers of gNBs with a fixed switch-to-gNB ratio. We compute the average latency gain by taking the average of pair-wise UE latency where each UE connects to a different gNB.}
    \label{fig:sim-latency-result}
    
\end{figure}

In this set of experiments, we evaluate the intra-cellular latency optimization behavior on larger topologies with multiple P4 switches using the Python simulation we described in Section~\ref{sec:implementation-multisw}. As shown in Figure~\ref{fig:sim-latency-result}, we can achieve latency gain between 20\% and 50\%, where the average is between 35\% and 40\%. The results show that the higher gNB/SW ratio leads to more stable latency gain with respect to the number of gNBs as the higher gNB/SW ratio leads to less SW, i.e., smaller networks. We see similar results yet with less latency gain because of the cases where the intra-cellular optimization is not advantageous as it is on a topology with two gNBs.

\makeatletter
\setlength{\@fptop}{0pt}
\makeatother

\chapter{CONCLUSION}\label{chapter:conclusion}

This thesis presents an SDN architecture for the next-generation cellular networks to achieve URLLC, explicitly targeting networks with a high frequency of intra-network communication. It utilizes the programmable data planes between cellular core and radio network, and it presents the concept of \textit{intra-cellular optimization}, that allows pre-authorized in-network devices to communicate without the requirement of cellular control signals. Further, this thesis introduces a novel control structure, Unified Control Plane (UCP), on top of the Ethernet Layer with an adapted version of UE information distribution link-state routing. We implement the architecture in this thesis on the P4 programming language for programmable data planes with a 5G implementation and a UE/RAN simulator. We further implement a simulation framework to evaluate the performance of our design in large-scale topologies on Python. We show that the concept of \textit{intra-cellular optimization} can achieve latency reduction up to 2x, whereas it can tremendously increase the network security and monitoring capabilities compared to traditional cellular networks while having a ten-millisecond level of control latency.

%\cite{*}
\bibliographystyle{styles/fbe_tez_v11}
\bibliography{references}

% \appendix
% \chapter[AN APPENDIX TITLE THAT IS LONG AND THEREFORE\\
% 	\hspace*{2.95cm} NEEDS MANUAL ADJUSTMENT IN LATEX CODE TO FIT\\
% 	\hspace*{2.95cm} PROPERLY IN TABLE OF CONTENTS]{AN APPENDIX TITLE THAT IS LONG AND THEREFORE NEEDS MANUAL ADJUSTMENT IN LATEX CODE TO FIT PROPERLY IN TABLE OF CONTENTS}
% \section{Future Implementation Directions}

% \section{P4 code}

% \section{Emulation Configuration}

% \section{Simulation Software}

% % \section{Future Implementation Directions}

% % In this section, we briefly discuss the future directions of our research, including possibilities of hardware implementation and further experimentation. Other than our evaluation focus, a few different factors can also demonstrate the capabilities of our design. One such metric is jitter, variations in the packet arrival time, and mainly caused by congestion and path changes. Our jitter evaluation using the architecture in Figure~\ref{fig:implementation} unfortunately does not give a realistic view since we use network emulation on VMs without any outside network interference. We aim to extend our work using hardware-P4 switches and connect to a multiuser 5G network to demonstrate its capabilities, including its jitter performance

\end{document}